\documentclass[namedreferences,hyperref,optionalrh]{spr-sola}
\usepackage{graphicx}        
\usepackage{color}           

\usepackage{url}
\usepackage{multirow}
\usepackage{tabularray}
\usepackage{booktabs}
\usepackage{longtable}
\usepackage{array}
\usepackage[table]{xcolor}
\usepackage{amssymb}        
\usepackage{breakurl}        
\usepackage{wasysym} 
 \usepackage{relsize}
 \usepackage{amssymb}
 \usepackage{setspace}
 \usepackage{graphicx}

\usepackage{booktabs}

\usepackage{xcolor}
\usepackage{hyperref}
\hypersetup{
    colorlinks,
    linkcolor={red!50!red},
    citecolor={red!50!red},
    urlcolor={red!80!red}
}

\newcommand{\aap}{{\it Astron. Astrophys.}}

\chardef\us=`\_

\begin{document}

\begin{frontmatter}
\title{Sunspot Observations in 1684\,--\,1702: \\John Flamsteed and Philippe de La Hire}

\author[addressref=aff1,corref,email={ned@geo.phys.spbu.ru}]{\inits{N.V.}\fnm{Nadezhda}~\snm{Zolotova}\orcid{0000-0002-0019-2415}
}\author[addressref={aff2},email={mikhail.vokhmianin@oulu.fi}]{\inits{M.V.}\fnm{Mikhail}~\lnm{Vokhmyanin}\orcid{https://orcid.org/0000-0002-4017-6233}}

\address[id=aff1]{St. Petersburg State University, Universitetskaya nab. 7/9, 198504 St. Petersburg, Russia}

\address[id=aff2]{Space Climate Group, Space Physics and Astronomy Research Unit, University of Oulu, Oulu, Finland}

\runningauthor{N.V.~Zolotova, M.V.~Vokhmyanin}
\runningtitle{Sunspot Observations in 1684\,--\,1702}

\begin{abstract}

In this work, we present an extensive review and detailed analysis of sunspot measurements, drawings, and engravings made by John Flamsteed and, mainly, by Philippe de La Hire during the Maunder minimum. All available information and contemporary knowledge about the sunspot nature are shown. The coordinates, areas, and numbers of sunspots and sunspot groups are reconstructed. Based on these observations, La Hire, Jean-Dominique Cassini, and his son Jacques Cassini regularly published results that shed light on the purpose of sunspot measurements and the scientific paradigm of that time. In particular, astronomers believed that sunspots were recurrent over decades. We compare the reconstructed time-latitude diagram with those obtained by \citeauthor{Spoerer1889} (Ueber die periodicitat der sonnenflecken seit dem Jahre 1618..., \citeyear{Spoerer1889}) and \citeauthor{1993A&A...276..549R} ({\aap} \textbf{276}, 549, \citeyear{1993A&A...276..549R}). The sidereal differential rotation rate is estimated, and its latitudinal profile is reconstructed. We also evaluate the fraction of sunspot groups that obey or violate Joy's law.

\end{abstract}
\end{frontmatter}

\section{Introduction}
     \label{S-Introduction} 

Nowadays, historical archival data are of great interest for solar \citep{2022ApJ...927..193C, 2022SoPh..297...79I, 2022ApJ...941..151H, 2022SoPh..297..127W, 2023ApJS..269...53E, 2023MNRAS.523.1809I, 2023SoPh..298..122H, 2023JSWSC..13...33H, 2024ApJ...968...65C, 2024ApJ...970L..31H, 2024MNRAS.528.6280H}, astronomical \citep{2021AN....342..675N}, and geophysical \citep{2023AN....34420078W, 2024GSDJ...11..504L} studies. Old observations also contribute significantly to the recalibration of sunspot indices \citep{2023SoPh..298...44C, 2024SoPh..299...45B}.

The Maunder Minimum (1645\,--\,1715), known as a representative Grand Solar Minimum, poses a challenge for theoretical solar models aiming to reproduce the disappearance of surface activity and the asymmetric butterfly diagram. The Parisian solar observations by Jean Picard, the father and son La Hire, and the Cassini family form the richest sunspot data set produced during the Maunder Minimum. 

To a large extent, relying on the Mémoires of the French Royal Academy of Sciences, \citet{Spoerer1889} compiled a list of sunspot latitudes for the period 1672\,--\,1713. The most extensive and valuable work was done by \citet{1993A&A...276..549R} to reconstruct sunspot latitudes and rotation profiles from the Parisian observations. Unfortunately, all the reconstructed data have been lost. Apart from spot coordinates, little is known about the numbers and areas of individual sunspots.

In this study, we present a detailed analysis of all the available sources: sunspot position measurements, sunspot drawings, sketches, and published engravings made by John Flamsteed at the Greenwich Observatory and by Philippe de La Hire at the Paris Observatory. To provide context and clarify the aim of the observations, we also analyze publications by Flamsteed, La Hire, and father and son Cassini, as well compare their observations with the German ones analyzed in \citet{2018AN....339..219N}. We reconstruct the number, area, latitude, longitude, and other parameters of sunspots and their groups. The reconstructed data are provided in the electronic supplementary materials. All processed sunspot images are available at \url{http://geo.phys.spbu.ru/~ned/History.html}.

\section{Methods}
     \label{S-Methods} 
     
The solar ephemerides required for the calculation of the heliographic grid orientation and sunspot positions are inferred using the French Planetary Theory VSOP87 \citep{1988A&A...202..309B, 1991aalg.book.....M}. The latitudes and longitudes of the sunspots were calculated by weighting each pixel of the sunspot according to its angular distance. Longitudes were measured from the zero meridian at Greenwich Noon on 1 January 1854 and rotated with a sidereal period of 25.38 days, as conventionally fixed by Carrington.

Here, the errors are the standard deviation in an obtained data set. Actual uncertainties may be larger. For example, an error of one second in time measurements results in an inaccuracy of 1\,--\,$1.5^{\circ}$ in the sunspot position at the center of the solar disk, and about $3^{\circ}$ for a spot located $60^{\circ}$ from the center \citep[for details]{2025SoPh..300...17Z}. 

To determine the rotation rate, we apply two approaches. The first is a commonly used method operating by subtracting the area-weighted longitude distances for the two consecutive days and dividing this quantity by the time elapsed between the two observations \citep{1984ApJ...283..373H}. In the second approach, the rotation rate is derived for the full observation interval, i.e. between distant observations \citep{1951MNRAS.111..413N}. This method is more accurate, but loses information on the sunspot movement in between. 

\section{Flamsteed, 5 May\,--\,8 July 1684}
\label{Flams}

\begin{figure}    
\centerline{\includegraphics[width=1\textwidth,clip=]{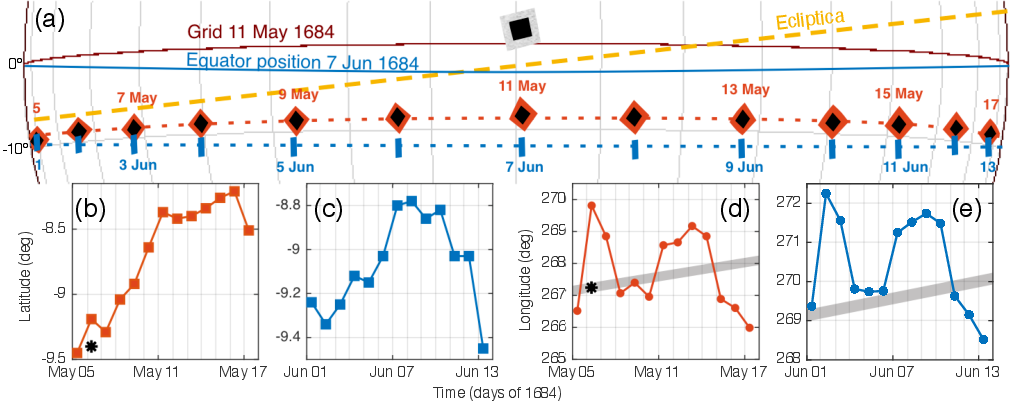}}
\small
        \caption{(\textbf{a}) Color-enhanced and $6^{\circ}$-rotated modification of the original engraving from \citet{1684RSPT...14..535F} with the heliographic grid and Gregorian dates imposed. The Equator position from 7 June is marked in blue. The black square on the top center is the original engraving of the sunspot scaled and oriented with the grid. (\textbf{b}) and (\textbf{c}) Sunspot latitudes deduced from the engraving. The latitude calculated by \citeauthor{1684RSPT...14..535F} on 6 May is shown with a \textit{black asterisk}. (\textbf{d}) and (\textbf{e}) The longitudes deduced from the engraving and that calculated by \citeauthor{1684RSPT...14..535F} (\textit{asterisk}). The \textit{gray shaded area} illustrates the expected longitudes at a rotation rate of $14.26^{\circ}$ d$^{-1}$.}
\label{Fig1}
\end{figure}

We begin with the series of sunspot observations made by John \citet{1684RSPT...14..535F}, the first Astronomer Royal, at Greenwich and published under the title ``An account of a spot seen in the Sun from the 25th of April to the 8th of May [Julian calendar] instant, with the line of its course predicted, if it make a second return." However, in contrast to the title, the engraving, accompanying the article, does not depict the actual sunspot tracks, but rather the expected path of the sunspot showing where it would appear every morning at 8 in May and June 1684.

\citeauthor{1684RSPT...14..535F} wrote that he saw a large sunspot while measuring the distance of Venus from the Sun, and it was the only spot he had seen since December 1676. \citeauthor{1684RSPT...14..535F} only referred to two actual observations on 5 and 6 May, and from the last one he inferred that the sunspot latitude was $-9.4^{\circ}$ and its central meridian distance was $66.75^{\circ}$ (according to measurements published in \citet{Flamsteed_1712}, it is $66.5^{\circ}$). He assumed that each point of the Sun rotated with a sidereal period of 25.25~days compared to 25.38~days used today for the heliographic grid. So, \citeauthor{1684RSPT...14..535F} himself defined the sunspot rotation rate as $14.26^{\circ}$ d$^{-1}$.

Figure~\ref{Fig1}a reprints the colored modification of the original engraving with an overlaid heliographical grid. \citeauthor{1684RSPT...14..535F} wrote: ``\textit{When the Spot was near the middle of the Sun it appeared very broad and almost square, the Nucleus of the same Figure about $40''$ diameter, shaped as it is designed... it seemed to have Consistence enough to endure a second return, if it shall it will enter the visible disk of the Sun on the 21 of May} [Julian calendar] \textit{in the evening...}"

On the top center of Figure~\ref{Fig1}a, we impose the original engraving of the sunspot scaled according to the size indicated in the text. Notably, in the engraving of the calculated sunspot track in May, the sunspot has a different orientation, and the diameter of $40''$ corresponds rather to the penumbra size, than to that of the umbra. This discrepancy crucially affects the penumbra/umbra areas, which are 804/214~msh according to the text note, or 334/72~msh according to the sunspot track on 11 May 1684.

Figure~\ref{Fig1}b and c shows the sunspot latitude extracted from the engraving of May and June. The latitude calculated by \citeauthor{1684RSPT...14..535F} on 6 May is shown by a black asterisk. The latitude increases almost gradually from $-9.5$ to $-8.3^{\circ}$ in May and from $-9.4$ to $-8.8^{\circ}$ in June. The latitude uncertainty is within $1^{\circ}$. Figure~\ref{Fig1}d and e shows the derived longitudes. The gray shaded area is bounded by two longitude values corresponding to the central meridian distances of $66.75^{\circ}$ and $66.5^{\circ}$ measured by Flamsteed, if the sunspot rotation rate is $14.26^{\circ}$ d$^{-1}$. Here, the discrepancy between the engraving and \citeauthor{1684RSPT...14..535F}'s calculations is up to $3^{\circ}$. More about uncertainties can be found in \citet{2025SoPh..300...17Z}.

\begin{figure}    
\centerline{\includegraphics[width=1\textwidth,clip=]{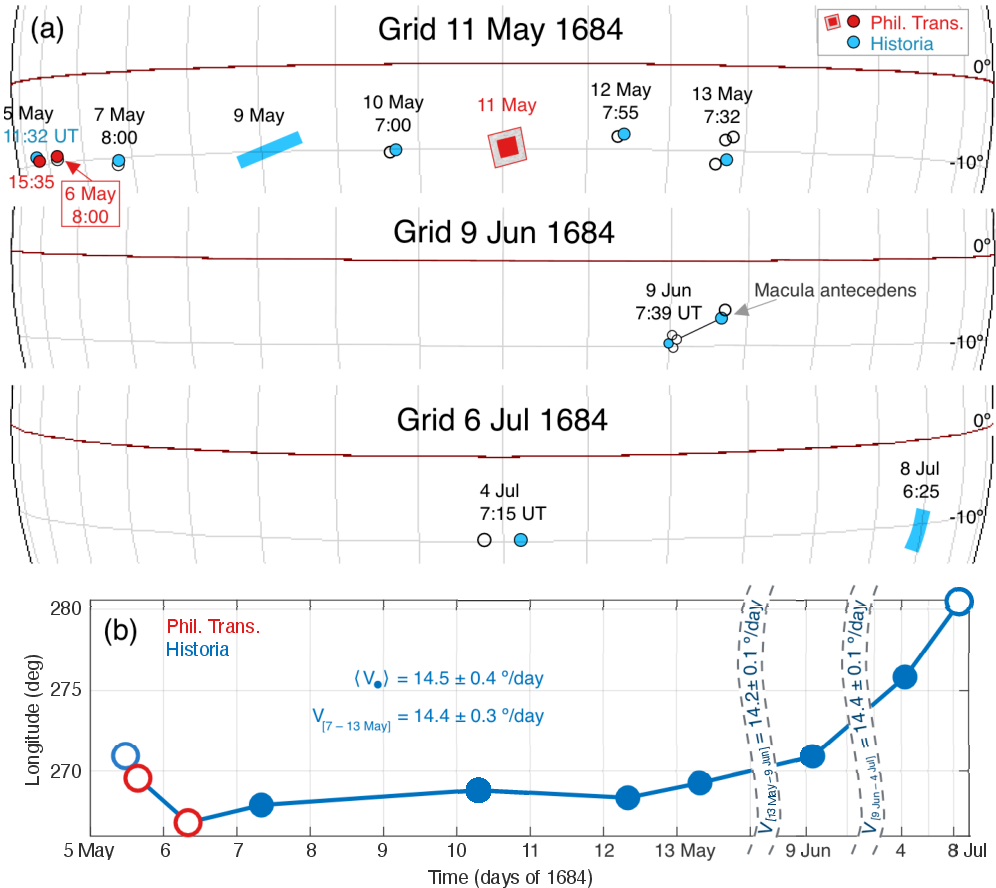}}
\small
        \caption{(\textbf{a}) Sunspot position derived from measurements in \citet{Flamsteed_1712} in \textit{blue} and \citet{1684RSPT...14..535F} in \textit{red}. Reliable data are represented by \textit{filled circles}, whereas unreliable data are shown as \textit{unfilled circles}. The \textit{blue rectangle} and \textit{annular sector} show the range of possible sunspot locations. The heliographic grid corresponds to the equatorial setup of a telescope. (\textbf{b}) Sunspot longitudes and rotation rate deduced based on reliable data.}
\label{Fig2}
\end{figure}

The actual sunspot measurements were published in \citet{Flamsteed_1712}. On 6 May 1684, Flamsteed recorded the solar diameter as $31'$ $36''$ using a 16-foot tube and $31'$ $28''$ using an 8-foot tube. The uncertainty in the solar diameter was apparently even greater, as in one measurement the limb-sunspot-limb distance was estimate to be $31'$ $14''$.

Figure~\ref{Fig2}a shows sunspot positions reconstructed from two publications: the Philosophical Transactions (in red) and from Historia (in blue). The size of the circles is arbitrary and does not reflect the actual size of the sunspots. In each observation, Flamsteed made a series of measurements, so we get several possible positions of a sunspot. Throughout this and the following figures, the filled circles mark subjectively more reliable data. The electronic supplementary material provides sunspot parameters reconstructed from both reliable and unreliable data. On 9 May, Flamsteed gave only the distance of the sunspot from the lower limb, thus only a possible range of sunspot locations is shown (blue rectangle aligned with the solar rotation axis). On 11 May, we map the original sunspot engraving with the umbra size of $40''$. 

On 9 June, Flamsteed noted the spot returning to the Sun again with a conspicuous ``companion'' about two minutes away from it and almost in the same declination parallel. He had not seen the ``companion'' in May, though he had been looking for it purposefully (\textit{Ante\`{a} n\`{o}n vidi etiamsi consulto quaesiveram}). However, on 6, 7, and 9 May 1684, \citeauthor{La_Hire_1683_1684} reported trailing sunspots of up to 209~msh, as we estimate. Thus, we suggest that Flamsteed was incapable of seeing non-prominent sunspots due to his observational approach and that the note about the absence of sunspots since 1676 should be taken with caution.

On 4 July, Flamsteed wrote that he saw the sunspot has returning again, meaning that Flamsteed believed that the active region survived over three rotations. On 8 July, only the distance between the sunspot and the limb is given. In Figure~\ref{Fig2}a, we show an apparent latitude-longitude range, where the center of the sunspot (\textit{Macula}) could have been located (blue annular sector). For the whole range of the observations (on 9 June, we use the trailing sunspot latitude), we evaluate the average latitude to be $-9.9 \pm 0.7^{\circ}$. Note that \citet{Spoerer1889} did not mention Flamsteed's observations.

Figure~\ref{Fig2}b presents the derived longitudes. Unfilled symbols indicate the near-limb observations, which we assume are less reliable for the rotation rate estimates. Based on the observations from 7 to 13 May, the average rotation rate (i.e., the average of all rotation rates between two adjacent observations) is $\left\langle V_{\bullet} \right\rangle = 14.5 \pm 0.4^{\circ}$ d$^{-1}$ in sidereal units. To reduce the impact of uncertainties, we also calculate the rotation rate between the distant measurements. The mean rotation rates are $V_{[7-13May]} = 14.4 \pm 0.3^{\circ}$~d$^{-1}$, $V_{[13May-9Jun]} = 14.2 \pm 0.1^{\circ}$~d$^{-1}$, and $V_{[9Jun-4Jul]} = 14.4 \pm 0.1^{\circ}$~d$^{-1}$. The fact that the longitudes of the object(s) match over three months suggests that it could have been the same recurrent active region or an activity nest rotating at $14.2 - 14.4^{\circ}$ d$^{-1}$.

\section{La Hire, Journal from 25 January 1683 to 10 October 1684}
\label{Tome1}

The core of this work is \citeauthor{La_Hire_1683_1684}'s hand-written observational journals stored in the Bibliothèque de l'Observatoire de Paris. The first 194-page journal was also published as a part of \citet{Monnier_1741}. We analyze both sources. In total, 313 measurements of the solar limb passages and 269 midday altitude measurements were reported. A typical observation provides the time between a solar limb and a midday altitude measurement. The observation days in La Hire's journals coincide with the spotless days according to the sunspot-group-number tables (based on La Hire's observations) in \citet{1998SoPh..179..189H} database. Presumably, \citeauthor{1998SoPh..179..189H} assumed that the observer also saw the passage of sunspots, if there were any on the disk, while measuring the passage of solar limbs.

The journal also provides measurements of the Sun's altitude in the morning and evening hours and notes on the accuracy of the pendulum. Solar observations are mixed with numerous reports on the passage of other celestial bodies: the Moon, Jupiter with its satellites, Saturn, Mercury, Venus, Mars, Sirius, Regulus, Arcturus, the Pole star, the brightest stars of Leo, and etc. The Sun was observed through a telescope equipped with a micrometer and the passage of a sunspot through a system of wires was measured by a pendulum and a chronicle. More about observational techniques and instruments can be found in \citet{1993A&A...276..549R} and \citet{ Boistel_2004}.

\begin{figure}    
\centerline{\includegraphics[width=1\textwidth,clip=]{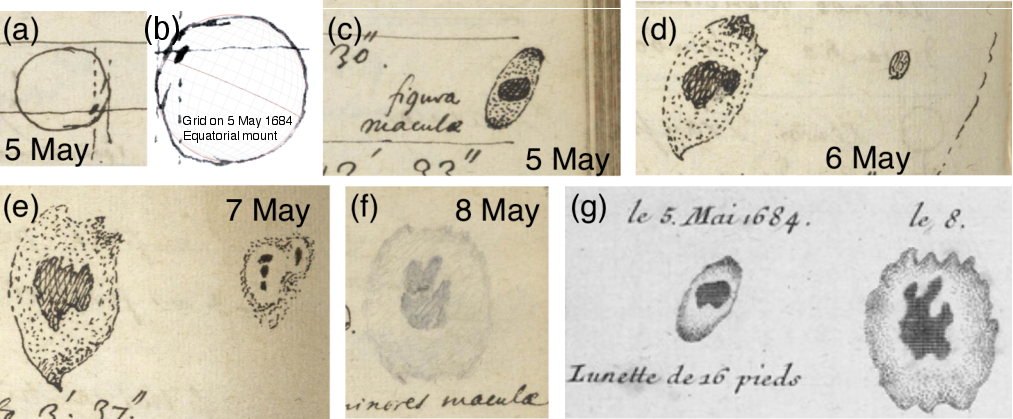}}
\small
        \caption{Sunspot drawings and engravings from 5 to 9(!) May 1684: (\textbf{a\,--\,f}) from \citet{La_Hire_1683_1684} and (\textbf{g}) \citet{Monnier_1741}.}
\label{Fig3}
\end{figure}

Figure~\ref{Fig3} illustrates several original sunspot images from the journal on scale. Figure~\ref{Fig3}a shows a small sketch of a sunspot on a poorly drawn solar disk. Figure~\ref{Fig3}b shows a modified image (resized, reversed, and flipped) with the heliographic grid corresponding to the equatorial mount of a telescope. According to the sketch, this sunspot appeared in the northern hemisphere, while the sunspot position measurements placed this active region in the opposite hemisphere. In this work, we therefore do not rely on such schematic drawings to reconstruct sunspot coordinates.

Figure~\ref{Fig3}c\,--\,f shows the sunspot group from 5 to 8 May (or supposedly 9 May 1684). The journal also contains a tracing paper from the image on 7 May, probably prepared for engraving. Figure~\ref{Fig3}g shows the same sunspot on 5 and 8 May engraved in \citet{Monnier_1741}. Drawings and engravings are somewhat different, hence they could have been based on other drawings. \citet{Monnier_1741} also contains sunspot drawings from 1 to 9 July 1684 differing slightly from those in \cite{La_Hire_1683_1684}.

\begin{figure}    
\centerline{\includegraphics[width=1\textwidth,clip=]{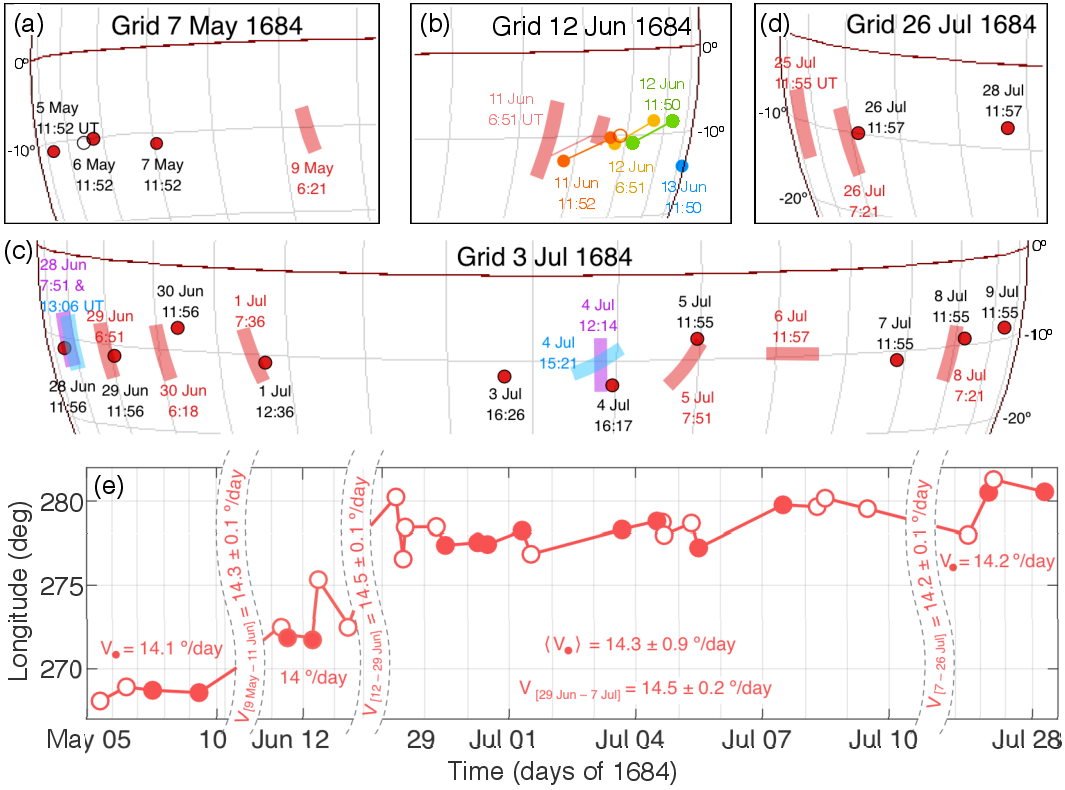}}
\small
        \caption{(\textbf{a}) Sunspot position derived from the measurements by \citet{La_Hire_1683_1684} from 5 May to 28 July 1684. Reliable data are represented by \textit{filled circles}, whereas unreliable and incomplete data are shown as \textit{unfilled circles} and annular sectors. The heliographic grids correspond to the equatorial mount. (\textbf{b}) Sunspot longitudes and rotation rate.}
\label{Fig4}
\end{figure}

Figure~\ref{Fig4}a,--,d shows the sunspot positions aligned with the rotation axis according to the measurements made by \citet{La_Hire_1683_1684} on 5 May to 28 July 1684.

There are two measurements at midday on 6 May that determine the sunspot position in the horizontal plane, giving two possible sunspot locations. We mark a more reliable position with a filled circle. The electronic supplementary material includes two versions of the derived sunspot parameters.

On 11\,--\,12 June 1684, the sunspot group consists of leading and trailing parts, which we schematically illustrate with linked circles (Figure~\ref{Fig4}a).

For the entire time interval, midday observations yield an average sunspot latitude of $-9.8 \pm 1.4^{\circ}$. Note that referring to Cassini's publications in Mémoires, \citet{Spoerer1889} gave the following estimates for the latitude of this sunspot: $-11^{\circ}$ from 5 to 17 May, $-10.8^{\circ}$ from 28 June to 9 July, and $-9^{\circ}$ from 26 to 28 July 1684.

Figure~\ref{Fig4}e illustrates the derived longitudes. Based on the most reliable observations (filled circles, which were usually made at midday, and when the sunspot was far from the limb), we determine that the rotation rate was $14.1 \pm 1^{\circ}$ d$^{-1}$ between 7 and 9 May, $14 \pm 1^{\circ}$ d$^{-1}$ between 11 and 12 June, and $14.2 \pm 1^{\circ}$ d$^{-1}$ between 26 and 28 July. The uncertainty is large for such short time measurements of one or two days apart. The average rotation rate between adjacent observations from 29 June to 7 July is $\left\langle V_{\bullet} \right\rangle = 14.3 \pm 0.9^{\circ}$ d$^{-1}$ and the mean rotation rate is $V_{[29Jun-7Jul]} = 14.5 \pm 0.2^{\circ}$ d$^{-1}$.

Similarly to Section~\ref{Flams}, we also calculate the rotation rate assuming a recurrent activity nest. The mean rotation rates are determined to be $V_{[9May-11Jun]} = 14.3 \pm 0.1^{\circ}$ d$^{-1}$, $V_{[12-29Jun]} = 14.5 \pm 0.1^{\circ}$~d$^{-1}$, and $V_{[7-26Jul]} = 14.2 \pm 0.1^{\circ}$ d$^{-1}$. Overall, in Figure~\ref{Fig4}, the longitude gradually increases, indicating that the active region rotated faster than the sidereal Carrington rotation at $14.2 - 14.5^{\circ}$~d$^{-1}$ aligning well with the measurements by Flamsteed. 

\begin{figure}    
\centerline{\includegraphics[width=1\textwidth,clip=]{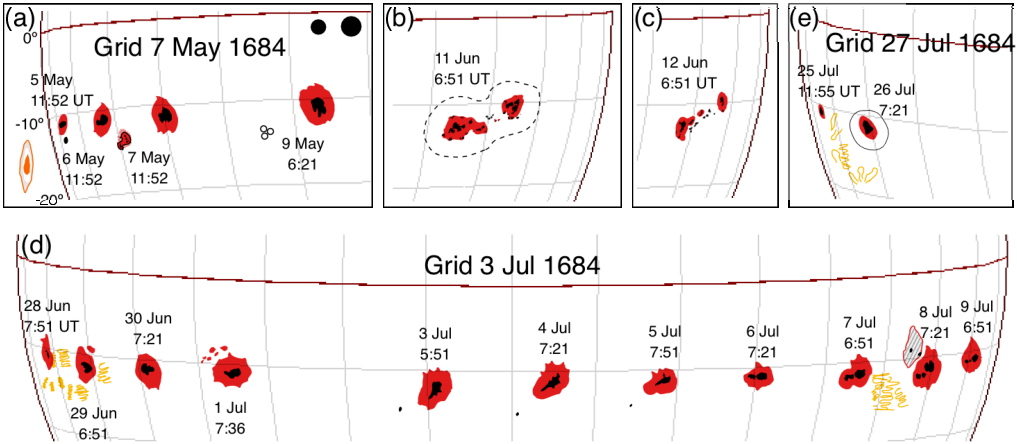}}
\small
        \caption{Visualization of sunspot transits reconstructed from the  measurements and sunspot drawings by \citet{La_Hire_1683_1684}. Yellow denotes faculae. Black circles indicate objects with the size of $30''$ and $40''$. The orange sunspot out of the solar disk reproduces the horizontally and vertically mirrored sunspot engraving from the Journal des Sçavans (\citeyear{Scavans_1684}).}
\label{Fig5}
\end{figure}

Observations in the morning hours in \citet{La_Hire_1683_1684} are usually accompanied by sunspot drawings. By reversing and flipping these drawings and combining them with the measurements, we present our assumption of the active regions track in Figure~\ref{Fig5}. The electronic supplementary material also includes the parameters derived from this reconstruction. These data should be considered with caution as little is known about the sunspot linear size. Our estimations rely solely on the drawings depicted together with the limb (Figure~\ref{Fig3}d, as an example), while toward the disk center, the size of sunspots is completely subjective. The uncertainty of our sunspot mapping is about $2^{\circ}$.

In the upper right corner of Figure~\ref{Fig5}a, we impose two circles of size $30''$ as noted in the Journal des Sçavans (\citeyear{Scavans_1684}), and $40''$, as noted in \citet{1684RSPT...14..535F} on 11 May 1684. An orange sunspot outside the solar disk is the sunspot engraving from the Journal des Sçavans (\citeyear{Scavans_1684}) on 5 May, mirrored horizontally and vertically, and scaled to a vertical umbral size of $40''$. The spot on the disk from the same day (also shown in Figure~\ref{Fig3}c) is scaled so that the horizontal size of its umbra is equal to the horizontal size of the sunspot outside the disk. Here, the uncertainty of sunspot area may reach a factor of three. This sunspot was reported as a new (\textit{la nouvelle Tache}) in \citet{Monnier_1741}. 

The next day, a small trailing sunspot was registered in \citeauthor{La_Hire_1683_1684}'s journal, which grew in size on 7 May and consisted of four umbrae (Figure~\ref{Fig3}d and e). On 9 May (8 May in La Hire's note), the sunspot was reported to have minor ones (\textit{minores maculae}), as observed with a 16-foot tube (\textit{ope Tubi 16 ped.}). These trailing spots were not drawn, and we mark them with three unfilled circles. According to Flamsteed, the sunspot had no trailing ones in May, although he was purposefully looking for them.

On 11 June, in \citet{Monnier_1741}, the sunspot group was described as the two new spots which began to appear (\textit{2 nouvelles Taches qui ont commencé à paroître}). Based on the detailed pencil drawing in \citeauthor{La_Hire_1683_1684}'s journal, the sunspot group had a complex structure and was outlined by a dashed curve (Figure~\ref{Fig5}b). Apparently, this was a halo observed around sunspots approaching the limb. We count 11 umbrae in the leading segment, 15 umbrae in the trailing segment, and 6 umbrae in between. The next day, we identify 2, 5, and 9 umbrae, correspondingly (Figure~\ref{Fig5}c). We are able to determine the size of this sunspot group as it was depicted alongside the limb. The sunspot on 26 July was also outlined.

On 28 June, \citet{Monnier_1741} wrote that a new sunspot appeared on the Sun (\textit{Il a paru une nouvelle Tache dans le Soleil}). We schematically redraw faculae from \citeauthor{La_Hire_1683_1684}'s journal in yellow zigzags. Two sunspot drawings on 29 and 30 June were also reproduced in \citet[Figure~1, therein]{1993A&A...276..549R}. On 1 July, supposedly trailing sunspots violate the Joy's law. There is only one trailing pore from 3 to 5 July, so that the objects on 1 July could have been something else, but not sunspots. On 8 July, a sunspot-like object reappeared with two umbrae again violating the Joy's Law. La Hire had hatching on this object. The sunspot was reported to have gone beyond the limb on 10 July, and faculae took its previous place.

Both \citet{Monnier_1741} and \citeauthor{La_Hire_1683_1684} wrote about the sunspot return on 25 July (\textit{Retour de la Tache} in French and \textit{Macula Solaris rediit} in Latin) which disappeared on 29 July (\textit{Macula evanuit}).

Considering the variability of the fine structure of the leading spot from 5 May to 26 July 1684 we are inclined to believe that the reported active regions, located at about $-10^{\circ}$ in latitude, were not the same long-lived recurrent sunspot group, but rather an activity nest persisting over four rotations. We will discuss this in more detail in the next section.

\section{Sçavans, Mémoires, and German Observations}
\label{Scavans}

\begin{figure}    
\centerline{\includegraphics[width=1\textwidth,clip=]{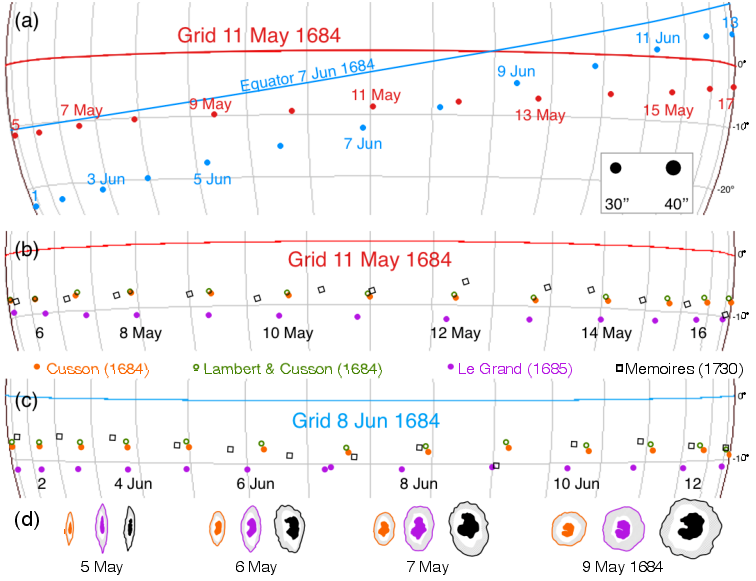}}
\small
        \caption{(\textbf{a}) Color-enhanced and rotated modification of the original engraving from the Journal des Sçavans (\citeyear{Scavans_1684}) with imposed heliographic grid. The Equator position from 7 June 1684 is marked in blue. Black circles indicate $30''$ and $40''$ objects. (\textbf{b}) and (\textbf{c}) Superposition of the sunspot transits from four engraving copies aligned with the heliographic Equator in May and June 1684: three editions of the Journal des Sçavans and one edition of Mémoires de l'Académie royale des sciences. (\textbf{d}) Horizontally and vertically reflected sunspot reproductions from various editions.}
\label{Fig6}
\end{figure}

An article in the Journal des Sçavans (\citeyear{Scavans_1684}), the earliest scientific journal published in Europe, reprinted in Mémoires de l'Académie royale des sciences (\citeyear{Memoires_1730}) was based on the Parisian observations in May 1684. We retell here this report because of its scientific value. The author (presumably Cassini) referred to the actual sunspot measurement on 5 May and predicted that, according to his theory of sunspot motion, it would pass at a distance of one and a half arcminutes from the disk centre by noon on 11 May. We roughly estimate that this corresponds to a longitude of $270.5^{\circ}$, whereas the actual measurements on 5\,--\,9 May (Figure~\ref{Fig4}) yield a longitude of $268.6 \pm 0.3^{\circ}$.

The author further predicted the sunspot passage in June in order to test the theory and, if necessary, correct it. As noted, tracks in May and June diverge because of the difference in the orientation of the Sun. The author said that he was waiting for the reappearance of the sunspot. He also analyzed the forshortening effect, which can be different on the two limbs because of the appearance of new sunspots and the decay of old sunspots. He noted that so far only the figure, size, color, motion, formation, physical changes, and dissipation of sunspots have been known, while their nature and origins have remained still hidden. 

We have found three copies of the Journal des Sçavans published by Cusson, Lambert and Cusson, and Le Grand. Figure~\ref{Fig6}a shows the color-enhanced and rotated modification of the original engraving by Cusson. We impose the heliographic grid of 11 May and the Equator position of 7 June 1684 as they would be seen through a telescope with an equatorial mount. Both predicted sunspot tracks are inclined with respect to the solar Equator. The slope varies from 4 to $7^{\circ}$, depending on the edition, which implies that the tracks could have been aligned with the Ecliptic. On the contrary, the tracks were described in the text as if they were oriented through an equatorial-mounted telescope. The engraving in June, reprinted in Mémoires de l'Académie royale des sciences (\citeyear{Memoires_1730}), yields even a bigger angle between the sunspot track and the solar Equator ($14^{\circ}$). No comments on the orientation of the sunspot tracks were provided.

Figure~\ref{Fig6}b and c show the tracks aligned with the solar Equator. The discrepancy between different copies is prominent, so we decide not to reconstruct the sunspot coordinates.

The sunspot on 5, 6, 7, and 9 May 1684 was also engraved on the border of the solar disk. Figure~\ref{Fig6}d represents sunspots from three editions keeping their proportions. The author wrote that the sunspot underwent various changes from its first appearance and was accompanied by other small spots as well as several faculae.

The difference in the fine structure of this sunspot can be seen when we compare these engravings with the drawings from La Hire's journal (Figure~\ref{Fig3}), and with the engraving from \citet{Monnier_1741}. The umbra appears to have been rotated by $90^{\circ}$ compared to Figure~\ref{Fig3}. Note that sunspot drawings are sometimes rotated by $90^{\circ}$ for various reasons \citep{2021SoPh..296....4V}. This sunspot was independently observed and drawn by Giovanni Domenico Cassini.

The author of the Journal des Sçavans had no doubt that the diameter of this sunspot including its nebula (``\textit{nebulosite}", i.e. penumbra; sometimes the penumbra also called the atmosphere) was greater than that of the Earth and its (we assume umbra) apparent diameter exceeded half an arcminute. This description aligns well with that of \citet{1684RSPT...14..535F}. For illustrative purposes, in Figure~\ref{Fig6}a, we add objects of $30''$ and $40''$.

The author wrote that in the three-foot telescope, through which the spot was discovered, only a slightly elongated blackness was seen, but through a larger telescope this blackness was seen in the form of an oval-shaped nebulosity, five times as long as it was wide. It was ``\textit{like a gondola loaded with the spot, or like the ring of Saturn, to which the spot served as a globe}". The engraving on 5 May, published by Cusson (Figure~\ref{Fig6}d), best corresponds to the indicated proportions. Therefore, in Figure~\ref{Fig5}a, it was compared to the sunspot track reconstruction. This information helped to improve the sunspot area estimate.

Another report in Mémoires, apparently by \citeauthor{Cassini_Memoires_1730_Facula} (since \citeauthor{La_Hire_1683_1684} did not observe on 17 May) says that this sunspot was expected on the eastern limb at 6 in the morning on 1 June. However, Cassini found only a facula accompanied by three other smaller faculae similar to those with which it [the larger facula] went beyond the limb on 17 May. 

Further, the text is as follows: ``\textit{Usually sunspots are transformed into faculae, which remain for several days after the sunspot has completely disappeared. There was no doubt that these faculae were the remnant of an already transformed sunspot, since they appeared} [on 1 June] \textit{in the place, where the sunspot should have been, and no others were found on the rest of the Sun's surface... Its} [facula] \textit{distance from the limb} [on 1 June] \textit{was about the same as that which the sunspot had on 5 May at 2 o'clock in the afternoon; so that, supposing this facula to have been a remnant of a sunspot, its return to its former distance from the limb occurred in 27 days and two-thirds, instead of other sunspots which returned in 27 days and one-third.}", as we translate. \citeauthor{Cassini_Memoires_1730_Facula} further explained this irregularity within the framework of the theory mentioned earlier and added that sunspots have their own small motions like clouds.

We also would like to quote the following comment: ``\textit{After this spot turned into faculae, we did not expect to see it return to its first form, as such a case was never observed. Nevertheless, on 11 June, about 6 in the morning, it reappeared in the place where we calculated it would be... Two large spots} [bipolar sunspot group]... \textit{After determining their positions, it was found that the spot, farthest from the limb, was in the same place as the spot, which had appeared in May, and the spot closer to the limb} [the leading spot(s)] \textit{was new. These two spots were visible again on 12 June, and on the 13th only the old spot remained...}" Thus, we can conclude that Cassini described an activity nest and not the same sunspot as they believed.

After a new series of calculations, Parisian observers expected to see the sunspot in the eastern limb again on 27 June. They wrote that when the sunspot appeared, it had a penumbra, it was accompanied by faculae, and it was expected to be observed until 10 July 1684.

Together with other publications by Cassini, analyzed in \citet{2025SoPh..300...17Z}, these quotes reinforce our opinion that sunspot tracks and the theory of their motion (also mentioned by Flamsteed) were used for the development of a theory of the solar axis precession relative to the Ecliptic and other constellations (e.g. Virgo, Pisces, Sagittarius). Long-lived spots or those remaining at the same latitude were best suited for this task. The period of the Sun's rotation has also been a subject of much interest.

In this regard, we refer to the treatise ``New theory of sunspots" by Esprit Pézenas written and revised between 1766 and 1772 \citep{Boistel_2004}. Father Pézenas presented a geometric method for the calculation of the tilt of the Sun's Equator to the Ecliptic and the Sun's rotation period from several observations of the same sunspots. He referred to Jacques Cassini's ``Éléments d'Astronomie" that addressed the same task. Also, Father Pézenas warned that an error of $1'$ results in an error of $15^{\circ}$ in the movement of the spot, and hence one cannot be too precise or use too good instruments when observing.

Finally, we mention German observations of that period. Kirch and Ihle reported a remarkable sunspot on 6 May 1684, and a blank Sun on the previous day. \citet{2018AN....339..219N} analyzed two sunspot drawings and got an average latitude of $-20.6 \pm 3.5^{\circ}$. The sunspot was on the edge of the western limb and therefore would go beyond the limb the next day. They concluded that this spot may be a different object, as compared to the sunspot measured in Greenwich and Paris. Having the derived sunspot longitudes, we know that on 6 May the sunspot was near the eastern limb (Figure~\ref{Fig4}). By applying the horizontal inversion to Kirch's and Ihle's drawings, we find that all the observers reported the same conspicuous active region near the eastern limb. On the other hand, if the sunspot drawn by Ihle and Kirch was indeed located near the western limb, Flamsteed and La Hire would not have measured it, as its track across the disk was over and it was therefore unsuitable for their theory. However, the latter assumption is highly speculative.

On 28 June 1684, Kirch wrote the third entrance of a large sunspot which might possibly be visible a fourth time, as he mentioned. We compare the measurements on 5 and 6 July by Kirch and Schultz with those by La Hire and found that the latitudes and longitudes of the sunspots coincide within a couple of degrees.

Another German observer, Wurzelbaur, reported a bipolar sunspot group approaching the western limb on 11 and 12 June 1684. The sunspot longitudes obtained from Wurzelbaur's schematic drawing \citep{2023SoPh..298..113V} differ from those measured by Flamsteed and La Hire by $2$ to $7^{\circ}$. 

From 2 to 8 July, the sunspot drawn by Georg Eimmart had an average latitude of $-10.8 \pm 2.8^{\circ}$ and an average longitude of $279.6 \pm 3.2^{\circ}$. The area of Eimmart's sunspot was 200\,--\,250~msh, i.e., smaller than the penumbra, but larger than the umbra in Figure~\ref{Fig5}d. We find the sunspot, apparently umbra, area obtained from Eimmart's observation to be less reliable, because his drawing was schematic and likely did not reflect its actual size.

\section{La Hire, Journal from 12 October 1684 to 25 January 1686}
\label{Tome2}

This \citeauthor{La_Hire_1684_1686}'s journal was printed in \citet{Monnier_1741} almost in its entirety with 209 measurements of the solar limb passages and 180 midday altitude measurements, but some observations were still missing. Sunspot observations are not mentioned, and the dates of the notes from this journal coincide with spotless days listed in \citet{1998SoPh..179..189H} database. 

\section{La Hire, Journal from 26 January 1686 to 30 June 1687}
\label{Tome3}

\begin{figure}    
\centerline{\includegraphics[width=1\textwidth,clip=]{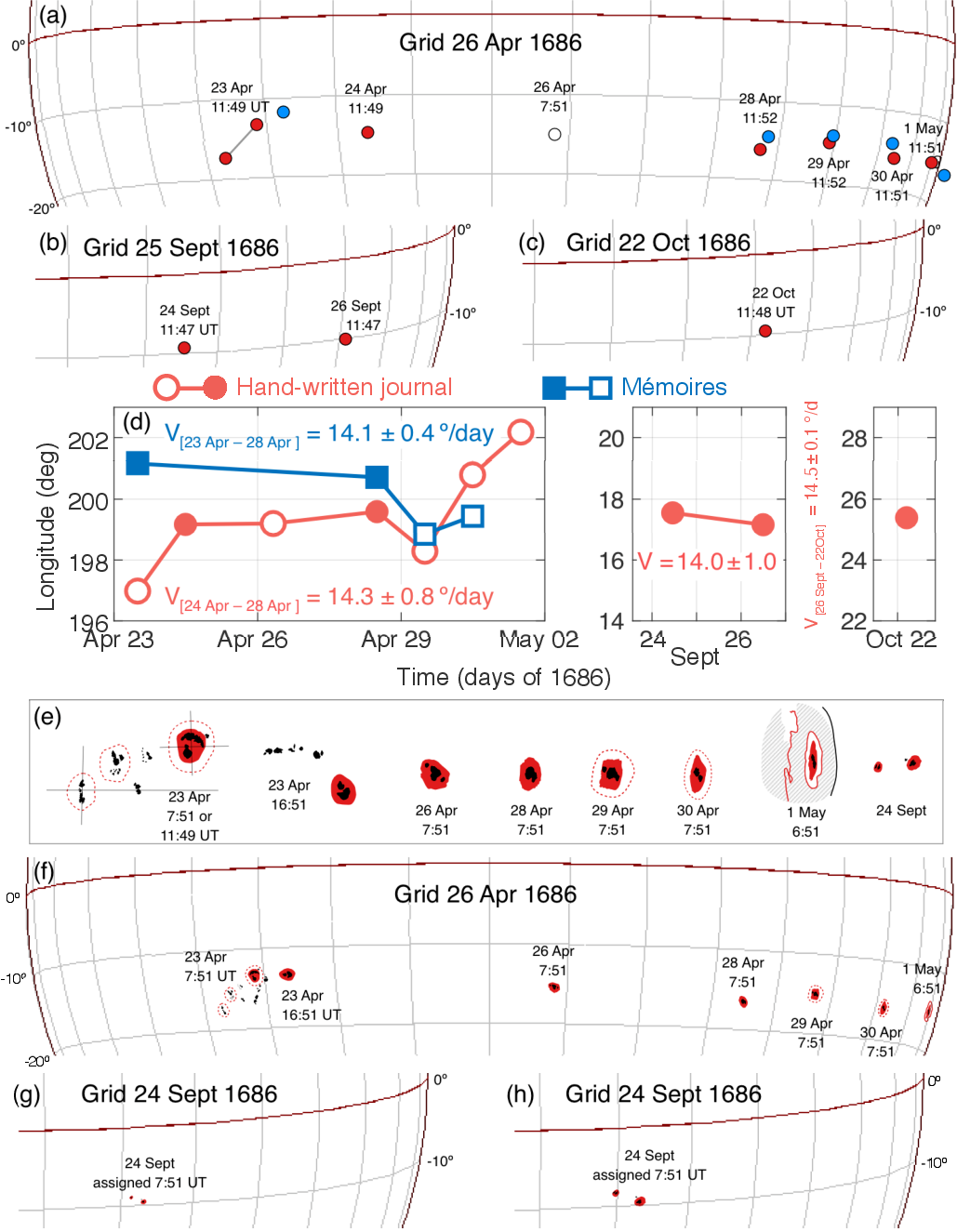}}
\small
        \caption{(\textbf{a\,--\,c}) Sunspot positions derived from the measurements by \citet{La_Hire_1686_1687} in \textit{red} and \citet{La_Hire_Memoires_1730} in \textit{blue}. Reliable data are represented by \textit{filled circles}. (\textbf{d}) Sunspot longitudes and rotation rate. (\textbf{e}) Horizontally and vertically reflected sunspots reproduced keeping their mutual scale. (\textbf{f})\,--\,(\textbf{h}) Reconstructed transit of the sunspot groups.}
\label{Fig7}
\end{figure}

This journal contains 283 measurements of the solar limb passages and 242 midday altitude measurements.

Figure~\ref{Fig7}a\,--\,c shows sunspot positions reconstructed from the measurements in \citeauthor{La_Hire_1686_1687}'s journal indicated by red circles of arbitrary size. On the morning of 26 April at 8 o'clock local time, the sunspot drawing was not accompanied by any measurements. The unfilled circle illustrates the possible position of this sunspot. 
The average latitude of the sunspot in spring is $-15.2 \pm 0.8^{\circ}$ and in autumn $-9.7^{\circ}$. \citet{Spoerer1889} provided an estimate of $-15^{\circ}$ using the data from \citet{La_Lande_1778} from 23 April to 1 May 1686.

Figure~\ref{Fig7}d shows the derived longitudes in red. On 23 April, the unfilled circle represents surprisingly a small longitude of the leading sunspot. To figure it out, we report a sequence of journal entries: at 8 in the morning (local time), there is a note saying that the previous day careful (\textit{diligenter}) observation had revealed no sunspots, and the today sunspot (23 April) was distant from the limb according to the observation. The note is followed by a sunspot drawing (\textit{``Figure Macula"}) and midday measurements. If we change the time from 11:49~UT to 7:51~UT (that is from 11:58 to 8:00 of local time), the longitude would be $199.16^{\circ}$ in agreement with the next observations. In April, we calculate the rotational rate only between 24 and 28 April and obtain $V_{[24-28Apr]} = 14.3 \pm 0.8^{\circ}$ d$^{-1}$, and 24\,--\,26 September, $ V_{[24-26Sept]} = 14.0 \pm 1.0^{\circ}$ d$^{-1}$. We also calculate the rotational rate between 26 September and 22 October and get $V_{[26Sept-22Oct]} = 14.5 \pm 0.2^{\circ}$ d$^{-1}$. Spring and autumn active regions are not from the same activity nest, otherwise their rotation rate would be $13^{\circ}$ d$^{-1}$.

Figure~\ref{Fig7}e illustrates horizontally and vertically mirrored sunspot reproductions. Their mutual size is preserved. The gray lines for the leading and trailing sunspots denote the lines to which the distance from the limb was measured by \citeauthor{La_Hire_1686_1687}. They allow us to find the size of the group on the first day. The most vague schematic drawing on 24 September 1686 was done with ink just in the corner of the page, like a small quick sketch. The other sunspots were drawn with a pencil in the middle of the pages. These detailed drawings were made with a larger telescope.

We have circled some sunspots with dashed lines corresponding to the lines in the drawings, but so thin that they are hardly recognized. On 23 April, the sunspot group was drawn twice. The orientation of the two images differs in the parallactic angle (the angle between celestial North and Zenith), implying that the sunspot group was observed in the evening with a telescope that was set at altitude-azimuth. 

By combining the individual sunspot drawings with the measurements, we present our assumption of the active regions tracks in Figure~\ref{Fig7}f\,--\,h. The most uncertain is the size of the sunspot group on 24 September: Figure~\ref{Fig7}g shows sunspots scaled as in Figure~\ref{Fig7}e, and Figure~\ref{Fig7}h is our subjective assumption with the leading sunspot enlarged to correspond to the size of the sunspots in April. We hope to clarify the size of this active region from Cassini's observations. The electronic supplementary materials includes the parameters derived from our reconstruction; these data should be considered with caution.

\section{Histoire, Mémoires, and German Observations}
\label{Histoire_1686}

Here, we give a rough summary of two scientific publications, by Cassini and La Hire, based on Histoire, Mémoires, and German observations. The journal Histoire~(\citeyear{Cassini_La_Hire_Hist_1686}) reports that the sunspot was in the middle of the Sun on 29 April at 8~p.m. following the parallel that deviated from the solar Equator by $27^{\circ}$ south. Several peculiarities can be noticed. The sunspot was in the middle of the disk on 25\,--\,26 April (Figure~\ref{Fig7}a). The observation could not have been made at 8 in the evening as the Sun was already below the horizon. Thus, it was most likely done at 8~a.m. as indicated in Cassini's hand-written journal. The $27^{\circ}$ distance should not be counted from the Equator, but from the horizontal line dividing the disk in two equal parts, when observing in a telescope with equatorial mount.

The report further states that: ``\textit{Mr. Cassini compared this spot with another one, which he observed on the same parallel in May 1684. It passed by the middle of the Sun on 11 May, 4 hours before Noon. Between these two times lies an interval of 714 days and 12 hours, during which this spot, presumably the same, must have made 26 revolutions of 27 days 11 hours and 32 minutes each. This gave Mr. Cassini a reason to search whether, among the great number of sunspots observed by Scheiner, there were any that could be considered to be the same; he found one observed in 1625, which was in the middle of the Sun on 16 May at 4 o'clock..., it had the same declination as} [the spot] \textit{in this year. Dividing the interval between these two observations, Mr. Cassini found that this spot must have made 810 revolutions in 27 days 11 hours and 31 minutes, as he found from his own observations at a shorter distance between them. This period may be taken as a measurement of the revolutions of the Sun relative to itself...}" Thus, we can conclude that Parisian astronomers paid particular attention to recurrent active regions.

In Mémoires, \citet{Cassini_Memoires_1730} reported the rotation period determined for the permanent spots on Jupiter (\textit{Taches permanentes de Jupiter}). \citet{La_Hire_Memoires_1730} reported few sunspot measurements and the following hypothesis of their nature: sunspots appear on the surface of the liquid Sun as a result of the emergence of pieces of other solid matter of irregular shape, which is inside the Sun and does not change. Large and small sunspots are the result of how high above the liquid surface a piece of this solid matter floats and how many of them join together. The simultaneous appearance of several spots in the remote places of the solar disk is the surfacing of several pieces of this matter. Here, we point out that since 1672 \citep{2025SoPh..300...17Z}, there have been no reports of two or more sunspot groups living at the same time. Hypothetically, several sunspot groups could be present on the Sun at the same time, but small spots were missed with a small telescope commonly used for routine observations.

\citeauthor{La_Hire_Memoires_1730} further wrote that there were many small particles (\textit{petites particules}, probably pores), homogeneous to the spots, mixed with all the matter of the Sun. Around the spots, these particles were held together (\textit{arrêter}) by the spots, so that the place around the spots appeared to be brighter (or cleaner). For this reason, when pieces of the solids submerged, faculae or luminous spots (\textit{facules ou des Taches lumineuses}) appeared in their place. Depending on how these bodies were positioned with each other and how they arose in the flow of solar matter, they moved faster or slower. This corresponds to the observed deviation of sunspots from a uniform motion.

As previously mentioned, Mémoires contains a few sunspot measurements made between 23 April and 1 May 1686. Unlike the hand-written journal, where the sunspot position was measured as a distance from the limb, here it was found relative to the disk centre. Also, Mémoires provides the angular size of the disk on 24 and 30 April, which coincides precisely with our reconstruction in Section~\ref{Tome3}.

In blue, Figure~\ref{Fig7}a shows the reconstructed sunspot position. On 1 May, the sunspot appeared outside the disk. We do not find any reasonable explanation for the discrepancy between the sunspot positions in the hand-written journal and those published in Mémoires. 

For Mémoires, we get the average latitude of $-13.6 \pm 0.8^{\circ}$. The derived longitudes are imposed in Figure~\ref{Fig7}d. The rotation rate between 23 and 28 April is $14.1 \pm 0.4^{\circ}$ d$^{-1}$, slightly slower than the Carrington grid rotation. Clearly, the measurements from the hand written journal (red) yield a faster rotation. By averaging the red and blue points in Figure~\ref{Fig7}d, the rotation rate is determined to be $14.25 \pm 0.1^{\circ}$ d$^{-1}$.

\citet{2018AN....339..219N} obtained two possible solutions for the sunspot latitude in Kirch's observation: $-13 \pm 5^{\circ}$ and $4 \pm 5^{\circ}$. Here, we find that the sunspot was known to be in the southern hemisphere. From German observations, we determine the longitude of 195\,--\,$197^{\circ}$, which is 2\,--\,$3^{\circ}$ less than that of the Parisian measurements.

\section{La Hire, Journal from 1 July 1687 to 30 April 1689}
\label{Tome4}

\begin{figure}    
\centerline{\includegraphics[width=1\textwidth,clip=]{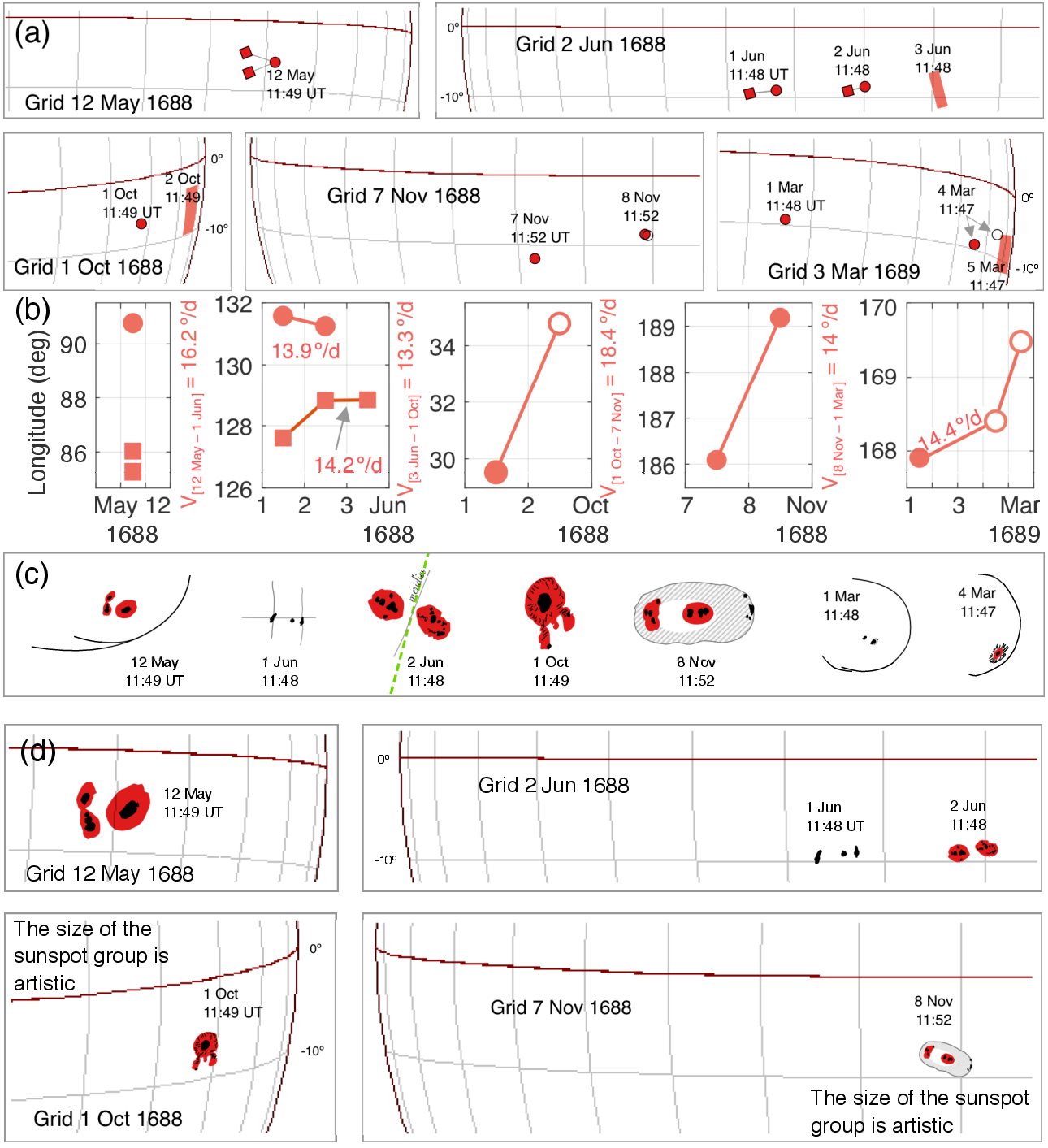}}
\small
        \caption{(\textbf{a}) Sunspot positions derived from the measurements by \citet{La_Hire_1687_1689}. Reliable data are represented by \textit{filled symbols}. \textit{Annular sectors} show the range of possible sunspot location. (\textbf{b}) Sunspot longitudes and rotation rate. (\textbf{c}) Horizontally and vertically mirrored sunspots reproduced by keeping their mutual scale. (\textbf{d}) Reconstructed transit of sunspot groups.}
\label{Fig8}
\end{figure}

This journal contains 352 measurements of the solar limb passages and 258 midday altitude measurements.

Figure~\ref{Fig8}a shows the reconstructed sunspot positions. Trailing spots are shown as squares. Figure~\ref{Fig8}b represents the longitudes. Filled symbols indicate sunspots with the central meridian distance less than $60^{\circ}$ which we assume to be more reliable. Due to the lack of long day-by-day observations, solely rotation-rate estimates between adjacent observations are not reliable. Active regions of November 1688 and March 1689 are possibly from the same activity nest rotating with $V_{[8Nov-1Mar]} = 14.0 \pm 0.03^{\circ}$ d$^{-1}$.

Note that Kirch reported a sunspot on 14 and 15 December 1688. \citet{2018AN....339..219N} reconstructed the latitude of this sunspot to be $-10.5 \pm 6.0^{\circ}$. We roughly estimate that its longitude was about $80^{\circ}$, i.e., different from the sunspots in November 1688 reported by \citeauthor{La_Hire_1687_1689}. Interestingly, \citeauthor{La_Hire_1687_1689} measured the limb passage on 9 and 17 December 1688, but did not report any sunspots. A possible reason is the use of a small 3-foot telescope in routine measurements at the Paris Observatory, which did not resolve small sunspots.

Figure~\ref{Fig8}c illustrates sunspot reproductions that were horizontally and vertically mirrored. Their mutual size is preserved. On 12 May 1688, there were two more blackish dots, which we do not reproduce here as they are different in color and look more like artifacts than pores. On 1 June, the gray lines are the lines to which the distance from the limb was measured. On 2 June, the gray oblique line representing the meridian drawn by \citeauthor{La_Hire_1687_1689} coincides with the actual meridian orientation indicated by the dashed green line. This confirms that the observer had drawn sunspots as they were seen through a telescope with an equatorial mount. Thus, the drawings provide fairly reliable information about the tilt-angle. On 1 October, the internal structure of the round penumbra was illustrated with a set of curves, apparently representing a fine penumbral texture. The three outgrowths do not differ in texture from penumbra, but form an atypical penumbra shape. Probably, they are not a penumbra, but rather faculae. These outgrowths also contain two small darker areas, which could be either real spots or an artifact from the pencil drawing. On 8 November 1688, the active region texture was quite complicated. This active region was observed far from the limb, thus the hatched region may not be a facula halo. On 1 March 1689, the sketch depicts a bipolar sunspot group oriented along the latitude (i.e., zero tilt angle), but due to the primitive nature of these sketches, this information is of little reliability. At midday on 5 March, the sunspot was reported to be almost touching the limb (``\textit{Macula limbum fere tangebat in meridie}").

Figure~\ref{Fig8}d represents our assumption on the active regions track. On 12 May 1688, measurements from \citeauthor{La_Hire_1687_1689}'s journal were incorporated with those from \citet{Cassini_Memoires_1730_1688} to get the size of the sunspot group. The latter appeared to have a large area of 1438~msh. However, on 8 and 10 May, \citeauthor{La_Hire_1687_1689} did not report sunspots while measuring the limb passages. The latitude of the sunspot on 12 May was $-13^{\circ}$ according to \citet{Spoerer1889}. We place the sunspot group closer to the Equator. The reason for the discrepancy is the approximate distance given by \citet{Cassini_Memoires_1730_1688} and used by \citeauthor{Spoerer1889}.

On 1 October and 8 November, we know nothing about the size of the sunspots, so the lower panels in Figure~\ref{Fig8}d show sunspots of arbitrary size. The orientation of the active regions is known, as they have been drawn through an equatorial-mounted telescope.

On 25 February and 2 March 1689, \citeauthor{La_Hire_1687_1689} measured the limb passages and did not report any sunspot, but drew sunspots on 1 and 4 March. This and other similar precedents are consistent with the point of view that solar observations, such as measuring the altitude of the Sun, its diameter or the limb passage, should not be considered equivalent to spotless days \citep{2021MNRAS.506..650H,2015SoPh..290.2719C,2016SoPh..291.2493C,2024MNRAS.528.6280H, 2025SoPh..300...17Z}.

In Mémoires, \citet{Cassini_Memoires_1730_1688} presented a new method to accurately determine the rotation of the Sun around its axis. No matter how hard the author tried to observe the Sun, he only succeeded in detecting sunspots on 12 May 1688 (originally 1686, but we think it is a typo). \citeauthor{Cassini_Memoires_1730_1688} used a small 3-foot telescope and at 6 in the morning saw two sunspots confirming the reported sunspots were large. 

These spots on 12 May 1688 appeared to Cassini in the same place on the solar disk, where he observed them on 14 May 1684 and 29 April 1686, so he counted their rotation period of 27 days 12 hours and 20 minutes. He also checked the observations by Scheiner and Hevelius to find sunspots at the same time of the year and at the same place. He found six candidates, with the largest time interval up to 836 solar rotations. From these temporally distant observations, Cassini concluded that the inaccuracy of the average period he calculated was of the order of 2 minutes. According to his observations, there were spots that returned several times to the same parallel of the Sun at certain times, and he was inclined to think that there were special places on the Sun suitable for sunspot formation.

\section{La Hire, Journal from 1 May 1689 to 10 October 1693}
\label{Tome5}

This \citeauthor{La_Hire_1689_1693}'s journal contains 767 measurements of the solar limb passages and 484 midday altitude measurements, but no sunspots were mentioned. Again, the dates of the journal coincide mostly with spotless days in \citeauthor{1998SoPh..179..189H} database. 

\section{La Hire, Journal from 14 October 1693 to 30 December 1696}
\label{Tome6}

\begin{figure}    
\centerline{\includegraphics[width=1\textwidth,clip=]{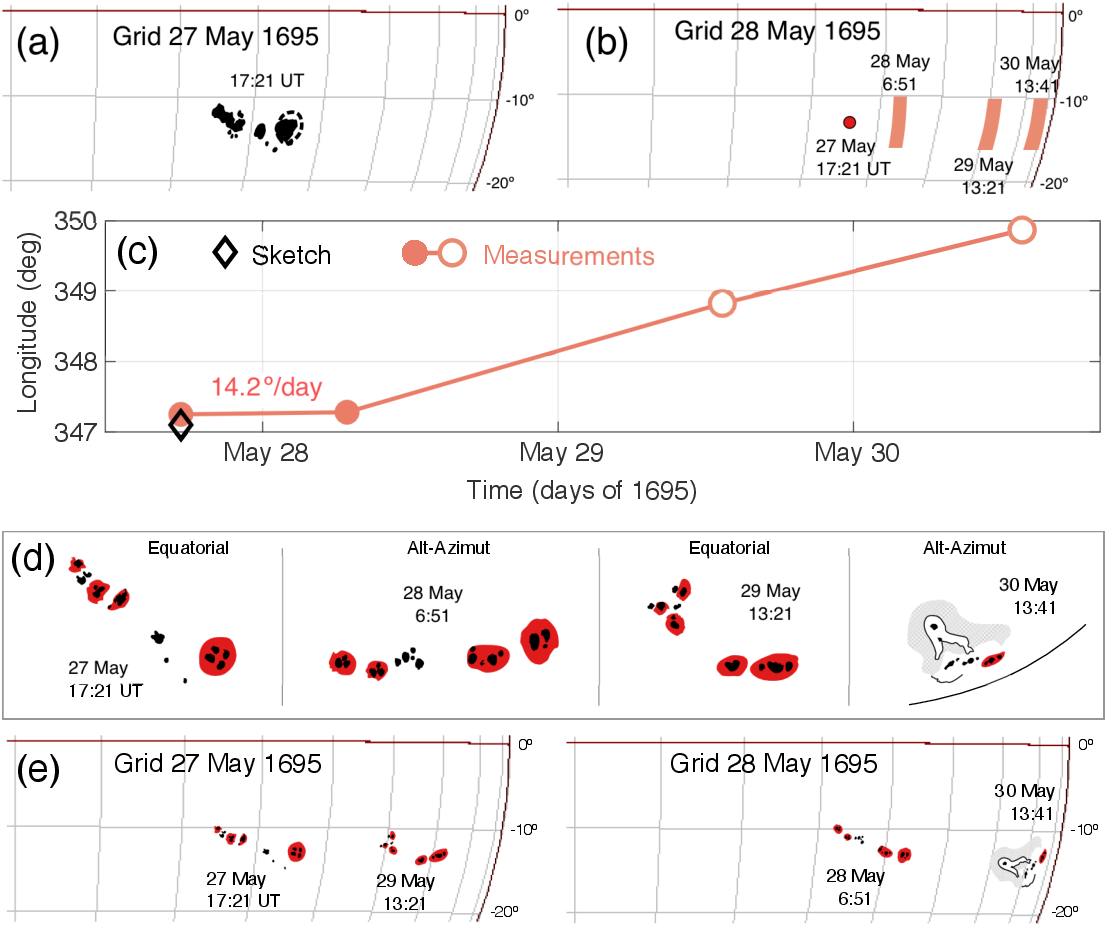}}
\small
        \caption{(\textbf{a}) Horizontally and vertically mirrored reproduction of the original sunspot sketch by \citet{La_Hire_1693_1696} with imposed heliographic grid. (\textbf{b}) Sunspot positions derived from the measurements. \textit{Annular sectors} show the range of possible sunspot locations. (\textbf{c}) Longitudes and rotation rate. Reliable data are represented by \textit{filled circles}. (\textbf{d}) Horizontally and vertically mirrored sunspots reproduced as they were seen with equatorial and alt-azimuth telescopes. (\textbf{e}) Reconstructed transit of the sunspot group.}
\label{Fig9}
\end{figure}

This journal contains 545 measurements of the solar limb passages and 494 midday altitude measurements. The angular size of the disk, which we use to determine sunspot coordinates, is $31'$ $41''$.

Figure~\ref{Fig9}a shows the horizontally and vertically mirrored sunspot sketch on the solar disk by \citet{La_Hire_1693_1696}, with the imposed heliographic grid. Since \citeauthor{La_Hire_1693_1696} did not measure the size of the sunspot group, we use this sketch to scale the size of the sunspot group in the individual drawings.

\citet{La_Hire_Hist_1695} wrote that no sunspots were observed since March 1689, and that it was a long time since any as large as those appeared. On 24 May 1695, there were none, and on 27 May at noon the sunspots were seen by \citeauthor{La_Hire_Hist_1695} for the first time.

Figure~\ref{Fig9}b illustrates the sunspot positions, reconstructed from \citeauthor{La_Hire_1693_1696}'s measurements. We do not fully understand the measurement system on 30 May, therefore, we use an incomplete data set. On 27 May, we estimate the latitude to be $-12.9^{\circ}$. \citet{Spoerer1889} yielded the latitude of $-12^{\circ}$ with a note that Cassini said that the great spot of May 1702 was in the same place.

Figure~\ref{Fig9}c represents the obtained longitudes. The rotation rate estimated from reliable adjacent observations is $14.2 \pm 1^{\circ}$~d$^{-1}$.

Figure~\ref{Fig9}d illustrates the horizontally and vertically mirrored sunspot reproductions. By analyzing the orientation of the sunspot groups, we conclude that two of them were taken with an equatorial-mounted telescope and two were taken with an altitude-azimuth mounted telescope. The boundaries of the penumbra were not clearly drawn, so their area may be evaluated with some uncertainty. On 30 May 1695, the white structure surrounded by a hatching does not look like a conventional penumbra. Figure~\ref{Fig9}e represents our reconstruction of the active-region track.

\section{La Hire, Journal from 4 January 1697 to 15 August 1704 (analyzed through 1702)}
\label{Tome7}

Solar observations dominate the celestial observations in this volume, with 1522 measurements of the solar limb passages and 1354 midday altitude measurements. We analyze sunspot observations up to 1702.

\subsection{November 1700\,--\,November 1701}
\label{Tome7_1}

\begin{figure}    
\centerline{\includegraphics[width=1\textwidth,clip=]{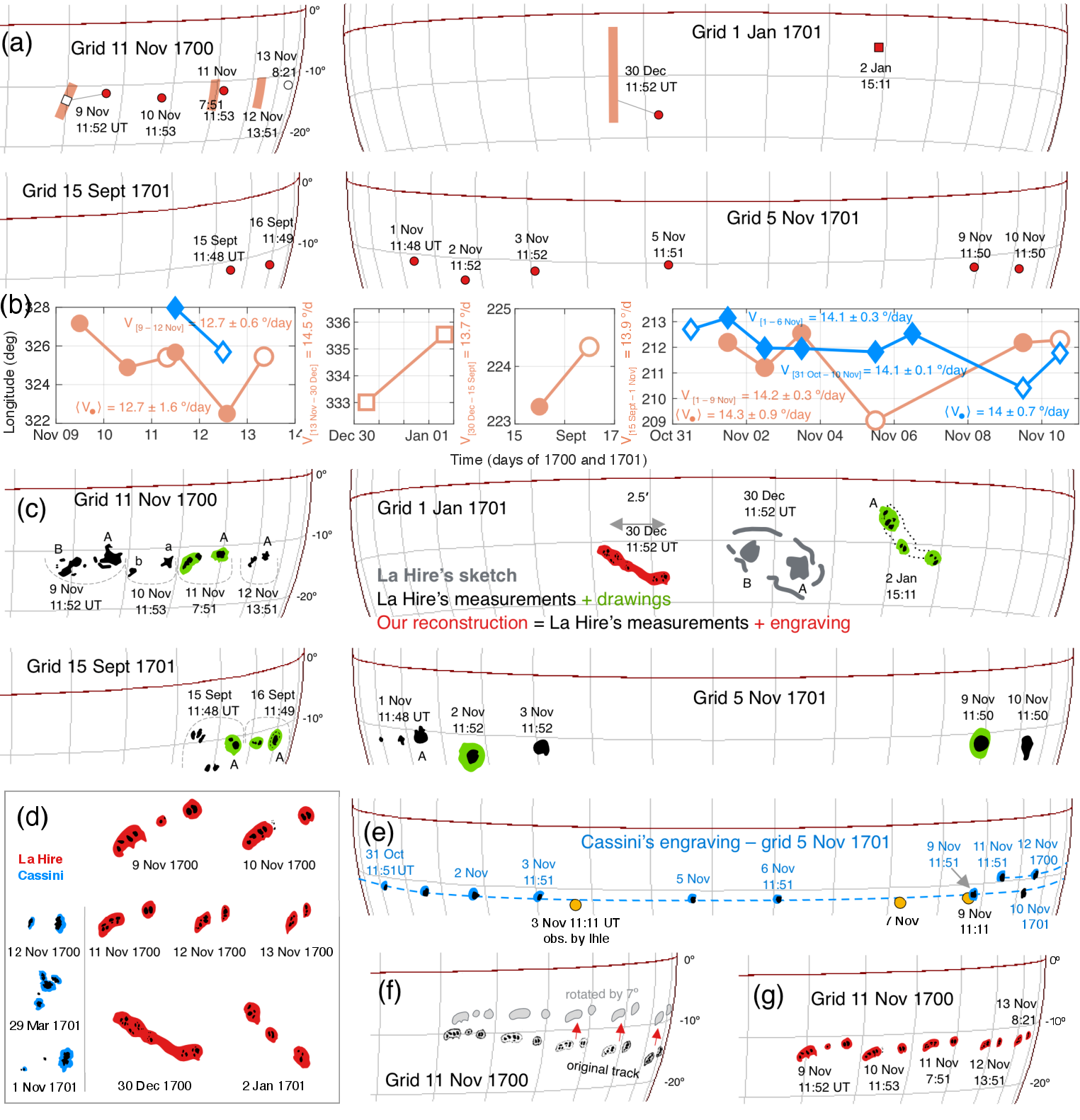}}
\small
        \caption{(\textbf{a}) Sunspot positions derived from the measurements by \citet{La_Hire_1697_1704}. (\textbf{b}) Longitudes and rotation rate: \citeauthor{La_Hire_1697_1704} in \textit{red} and \citeauthor{Cassini_1701} in \textit{blue}. Reliable data are represented by \textit{filled symbols}. (\textbf{c}) Sunspot tracks reconstructed from \citeauthor{La_Hire_1697_1704}'s measurements and drawings in \textit{black} and \textit{green}, one sketch in \textit{gray}, and our reconstruction in \textit{red}. (\textbf{d}) Horizontally and vertically mirrored sunspots from \citet{La_Hire_1701} and horizontally mirrored sunspots from \citet{Cassini_1701}. (\textbf{e}) Rotated and color-enhanced modification of the original engraving from \citet{Cassini_1701} with imposed heliographic grid. Sunspot positions reconstructed by \citet{2018AN....339..219N} are denoted by yellow circles. (\textbf{f}) Original engraving from \citet{La_Hire_1700} in \textit{black} and its rotated imprint in \textit{gray} with imposed heliographic grid. (\textbf{g}) Reconstructed transit of the sunspot group.}
\label{Fig10}
\end{figure}

Figure~\ref{Fig10}a illustrates the sunspot positions, reconstructed from the measurements in \citeauthor{La_Hire_1697_1704}'s hand-written journal. On 9 November 1700, the white square denotes the trailing sunspot position extracted from a sketch of the solar disk. On 13 November 1700, the white circle is our assumption on the sunspot location according to a short note that the sunspot is still on the Sun (\textit{macula ad huc appatebat}). On 16 September 1701, we correct the sunspot passage by one minute, and on 2 November 1701, we correct the sunspot altitude by $5''$ due to obvious typos. On 30 December and 2 January 1701, the sunspot latitude has large fluctuations, a square indicates the trailing sunspot position.

\citet{Spoerer1889} incorporated publications from Histoire and Mémoires de l'Académie Royale des Sciences and evaluated the sunspots latitude to be $-9.5^{\circ}$ on 7\,--\,13 November 1700 compared to our estimate of $-11.7 \pm 0.5^{\circ}$ for 9\,--\,11 November 1700. Then, $-3^{\circ}$ from 28 December to 2 January 1701, and $-12^{\circ}$ from 31 October to 10 November 1701 compared to our estimate of $-11.9 \pm 1.1^{\circ}$ for 1\,--\,10 November 1701.

Figure~\ref{Fig10}b shows in red the longitudes obtained from \citeauthor{La_Hire_1697_1704}'s measurements. Data from 9\,--\,12 November 1700 reveal a decrease in longitude with the average rotation rate between adjacent observations of $\left\langle V_{\bullet} \right\rangle = 12.7 \pm 1.6^{\circ}$~d$^{-1}$ and the rotation rate between temporally distant measurements $V_{[9-12Nov]} = 12.7 \pm 0.6^{\circ}$~d$^{-1}$. Such low values are not found in reliable observations, they are more likely due to measurement uncertainties. Data from 1\,--\,9 November 1701 yield $\left\langle V_{\bullet} \right\rangle = 14.3 \pm 0.9^{\circ}$~d$^{-1}$ and $V_{[1-9Nov]} = 14.2 \pm 0.3^{\circ}$~d$^{-1}$.

We also calculate the rotation rate between the series of the observations. We think that only the first and second observation series separated by two solar rotations may be related to the same activity nest. These observations correspond to the rotation rate of $V_{[13Nov-30Dec]} = 14.5 \pm 0.2^{\circ}$~d$^{-1}$. Note that according to the Greenwich catalogue, the longest-lived activity nest persisted for only nine rotations \citep{2010SoPh..262..299H}.

In black and green, Figure~\ref{Fig10}c illustrates sunspot positions reconstructed from the combination of measurements by \citeauthor{La_Hire_1697_1704} and sunspot sketches in his journal. These sunspots were drawn either on the solar disk or together with the solar limb, making it possible to estimate their size, though with limited accuracy. In contrast to the detailed pencil drawings in the previous journals, these sketches are mostly schematic and made with ink, yet still provide valuable information on sunspot-groups tilt-angle. Detailed individual sunspot engravings (Figure~\ref{Fig10}d, in red) were published by \citet{La_Hire_1700, La_Hire_1701} in Mémoires. These images were either drawn by another astronomer at the Paris observatory or represent additional drawings made by La Hire beyond those in his hand-written journal.

\citet{Cassini_Maraldi_1701} wrote that Wurzelbaur observed a sunspot from 7 to 13 November 1700. \citet{2018AN....339..219N} translated Wurzelbaur's letter to Kirch saying that the sunspot was already seen on 6 November in the afternoon. \citeauthor{La_Hire_1700} discovered the sunspots on 9 November 1700; due to bad weather, the previous observation was made on 31 October. He suggested that this sunspot could reappear on the eastern edge on 28 November. A short note on 21 December by \citet{La_Hire_1700_2} says: ``\textit{I could not see the Sun until two days after 28 November, on which day, according to my calculations, the sunspots... should have reappeared. I looked with great care to see if any sunspot or even faculae had reappeared, as we usually see after the disappearance of sunspots; but on the following days I could see nothing.}"

The two observations on 30 December 1700 and 2 January 1701 are the most uncertain. \citet{La_Hire_1701} wrote that he scrutinized the solar disk and did not observe sunspots on 28 December. The following days, except 30 December and 2 January, were cloudy. After half a rotation, \citet{La_Hire_1701} looked for the reappearance of these spots near the eastern limb, but saw nothing. On 11 and 22 January 1701, he noted that the spots have not reappeared (\textit{non apparent amplius macula}, from his observational journal).

The original sunspot sketch on 30 December is reproduced in gray (Figure~\ref{Fig10}c). The apparently enlarged image is mirrored horizontally and vertically and shows letters assigned to the sunspots by \citeauthor{La_Hire_1697_1704}. The longitude of the trailing sunspot does not match that measured on 2 January 1701. This discrepancy can be resolved by shifting the sketch to the eastern hemisphere at the same distance relative to the central meridian. We presume that \citeauthor{La_Hire_1697_1704} measured the sunspot position correctly, but made an incorrect sketch.

We combine La Hire's measurements with the sunspot engraving\footnote{This engraving by accident appears in Mémoires published in 1700. It definitely has to be on the same page, but in Mémoires published in 1701.} from \citet{La_Hire_1701}. The red blob represents the assumed mapping of the sunspot group in Figure~\ref{Fig10}c. The uncertainty of the sunspot sketch affected the text in \citet{La_Hire_1701}: all the measurements in Mémoires coincide with those in \citeauthor{La_Hire_1697_1704}'s hand-written journal, but they should be placed in the eastern hemisphere, not in the western (\textit{après le centre du Soleil}) hemisphere as \citet{La_Hire_1701} noted. The sunspot engraving (in red) on 30 December was additionally reduced to the size of $2.5'$, although \citeauthor{La_Hire_1701} pointed out that the size was determined imprecisely because of many small spots surrounding the large ones.

On 2 January 1701, \citeauthor{La_Hire_1697_1704} measured the closest distance between the umbra and the Equator (\textit{ad maculam inferiorem et negriorem}, marked with letter A). The fine structure of this sunspot group in the ink sketch (Figure~\ref{Fig10}c, 2 January) closely resembles that in the engraving (Figure~\ref{Fig10}d, 2 January).

Later in the chronology, \citet{Cassini_Maraldi_1701} and \citet{Cassini_1701}\footnote{The author of the article \citet{Cassini_1701} is labeled ``Cassini le fils", so we assume that Jacques Cassini, Giovanni Domenico Cassini's son, is the author.} reported a sunspot on 29 March 1701, the next day the Sun was not clearly visible, and on 31 March the complete disappearance of sunspots was noted. From the engraving in \citet{Cassini_1701}, we roughly estimate the sunspot latitude to be $-11^{\circ}$ ($-12^{\circ}$, according to the text) and the longitude about $251^{\circ}$. The text mentions three or four sunspots, with the largest one being round. Figure~\ref{Fig10}d reproduces in blue the individual engraving of the sunspot group on 29 May showing four sunspots and one pore. On the contrary, the corresponding engraving of the solar disk contains only one small sunspot and may not reflect the actual sunspot size.

\citeauthor{La_Hire_1697_1704} took solar measurements every day from 27 to 31 March 1701 and did not report any sunspots. Presumably, La Hire was making routine solar altitude measurements with a 3-fool telescope \citep{Cassini_1671_Ph_Tr, Cassini_Memoires_1730_1688}, which did not resolve small sunspots.

We do not reconstruct sunspot parameters on 29 March 1671 because the observation was made by Cassini, not by La Hire. This observation will be analyzed alongside the data from Cassini's journals in a future study.

On 14 September 1701, \citeauthor{La_Hire_1697_1704} measured the passage of limbs and the altitude of the Sun, but did not note any sunspots. The next two days, he measured and sketched sunspots. On 17 September, the sunspot was reported to have disappeared from the disk (\textit{Macula transiuit ad partes poster $\mathlarger{\mathlarger{\astrosun}}$}). We have found no mentions of these observations in Histoire and Mémoires.

The fourth series of observations was made in November 1701. On 31 October, \citeauthor{La_Hire_1697_1704} took the measurements, but did not report any sunspot. Note that he apparently used a small telescope which was unable to resolve the sunspot near the limb due to insufficient contrast \citep[for details]{1991QJRAS..32...35S, 1993ApJ...411..909S}. Moreover, \citeauthor{La_Hire_1697_1704} took measurements on 4 and 6 November and did not report sunspots, but on 11 November, he wrote that the spot may no longer be visible because of clouds on the adjacent limb (\textit{non apparet amplius macula forsitam propter nubes cum sirpropier limbo $\mathlarger{\mathlarger{\astrosun}}$}).

In his journal, \citeauthor{La_Hire_1697_1704}'s sunspot drawings made with ink were schematic. Nevertheless, on 1 November 1701, he drew trailing sunspots, which obey the Joy's law (Figure~\ref{Fig10}c, 1 November 1701). The area of the biggest umbra was 426~msh. Figure~\ref{Fig10}e illustrates in blue the engraving from \citet{Cassini_1701} rotated by $8^{\circ}$ with the imposed heliographic grid. The sunspots in November 1700 and 1701 are engraved with an average area of $158 \pm 40$~msh and without trailing spots; this differs substantially from \citeauthor{La_Hire_1697_1704}'s drawings in Figure~\ref{Fig10}c.

We translate \citet{Cassini_1701} as follows: \textit{we observed the spots at Rode on 11 November 1700 in the afternoon, when we took the altitude of the Sun to verify our clocks, and saw two long figures} [bipolar sunspot group], \textit{as they typically are, the largest of which was closer to the western limb... The next day, we observed at Noon... On 13 November} [1700], \textit{the weather was bad... On 29 March this year} [1701], \textit{while in Montpellier, we discovered other sunspots... 31 October} [1701]\textit{, I saw a sunspot... With a 40-foot telescope, I saw a single spot surrounded with penumbra} [\textit{entourée d'un Atmosphere}] \textit{and numerous faculae. On 1 November at Noon, we determined its position, observing it with a 40-foot telescope, we found four other spots of much smaller size located between the main spot and the eastern limb. On 2 November, one of the small spots disappeared and the other became a double spot. On 3 November, the small sunspots disappeared, and the main sunspot seemed to consist of two irregularly shaped spots joined together. On 4 November, at the time of the Sun's passage through the meridian, it was cloudy. On 5 November, the spot went through the vertical in a second. On the 7th and 8th, the sky was cloudy. On 9 November, the spot split into two unequal in size, the smaller of which was located to the South. On the 10th, the spots were observed without change in shape. On the 11th at Noon, the sky was not very bright and the spots were not visible. At that time, they were very close to the limb, which is a situation, when they are very difficult to observe.}

We point out ``\textit{as they typically are}'', however, historical astronomers rarely drew and engraved sunspots with trailing sunspots. Now we know that the sunspot group had trailing spots. The size of one second (in time) is fairly close to the size of the engraved umbra on 5 November 1701 (37~msh, with the entire sunspot area of 91~msh, Figure~\ref{Fig10}e). If we estimate the time uncertainty also to be one second, then the sunspot area uncertainty would be a factor of two near the disk center. In Figure~\ref{Fig10}c, \citeauthor{La_Hire_1697_1704}'s raw sunspot sketches in November 1701 apparently overestimate the sunspot area.

Comparison of the sunspot engravings and drawings on 12 November 1700 and 1 November 1701 by La Hire and Cassini (Figure~\ref{Fig10}c and d on the corresponding dates) reveals some discrepancies. According to \citeauthor{La_Hire_1701}, in November 1700 the sunspot followed Joy's law, while from \citeauthor{Cassini_1701}'s engravings, after a clockwise rotation by $24.8^{\circ}$ (the value of the positional angle on that day; we add a rotation to the individual sunspot engravings because they were oriented in accordance with sunspot appearance in a telescope), the sunspot group yielded zero tilt-angle or even slightly opposite to Joy's law (we will return to this issue when we will consider Cassini's journals). Comparison of the blue penumbra orientation in Figure~\ref{Fig10}d~and~e on 12 November 1700 also reveals the discrepancy. Eventually, we are inclined to trust more \citeauthor{La_Hire_1697_1704}'s day-by-day observations showing the proper tilt-angle.

On 12 November 1700\footnote{Originally dated 12 November 1701, this is likely a typo.}, \citet{Cassini_1701} estimated the central meridian distance of the sunspot to be $66^{\circ}$. We get almost the same value of $66.26^{\circ}$ indicating that this engraving (Figure~\ref{Fig10}e) has been made with great accuracy and is suitable for estimating the sunspot coordinates.  Figure~\ref{Fig10}b shows the reconstructed longitudes of this sunspot in blue. Utilizing only reliable data, we get the average rotation rate of $\left\langle V_{\bullet} \right\rangle = 14.0 \pm 0.7^{\circ}$~d$^{-1}$ and $V_{[1-6Nov]} = 14.1 \pm 0.3^{\circ}$~d$^{-1}$, while for the entire interval $V_{[31Oct-10Nov]} = 14.1 \pm 0.1^{\circ}$~d$^{-1}$. 

\citeauthor{Cassini_1701} used a synodic sunspot rotation rate of $13.1^{\circ}$~d$^{-1}$ (that is $14.1^{\circ}$~d$^{-1}$ in sidereal units) calculated in other observations and estimated that the sunspot crossed the central meridian on 7 November 1700 near noon and on 29 March 1701 at 8 in the evening. Thus, we are delighted to mention that Cassini's rotation rate is close to our estimates.

In black, Figure~\ref{Fig10}f displays the original engraving of the solar disk from \citet{La_Hire_1700}. Since \citeauthor{La_Hire_1697_1704} used a telescope with an equatorial mount, we are able to determine the orientation of the heliographic grid. The original sunspot track goes down, implying that \citeauthor{La_Hire_1700} aligned sunspots with the ecliptic. The corrected imprint rotated by $7^{\circ}$ is shown in gray. The sunspot latitude from the gray imprint deviates from those in measurements and the longitudes exhibit a wider scatter. Thus, this engraving is not suitable for the reconstruction of sunspot parameters. In Figure~\ref{Fig10}g, we combine measurements in Figure~\ref{Fig10}a and individual sunspot engravings. The corresponding sunspot parameters together with other recovered data are included in the electronic supplementary materials.

We now summarize the conclusions made by the Parisian astronomers on the series of observations published in Histoire and Mémoires. \citet{La_Hire_1700} wrote about the active region in November 1700 that this sunspot had not been seen since May 1695, when it had more or less the same shape, but one cannot be sure that it was the same sunspot. He reminded his hypothesis of irregularly shaped solid matter showing itself from different sides from time to time, and revolving in liquid solar matter with uniform motion. He concluded that 73 rotations had elapsed between these observations, resulting in a sunspot rotation period, or the Sun's rotation around its axis, of 27 days 7 hours 7 minutes.

\citet{Cassini_1700} wrote approximately the following: ``\textit{without sunspots, we would not be able to determine the rotation period of a uniformly luminous star. Sunspots often disappear for a considerable time, then, as soon as they reappear, all attention is again drawn to them to check whether the period of 27 days is accurately determined. When after several years during which the spots have not been seen, we begin to see them again, we generally see them at the time and place in the Sun where they should appear, owing to the period of 27 days, which proves that spots do not drift through the body of the Sun, but are fixed in a definite place}". 

Further, \citet{Cassini_1700} wrote that he had never seen sunspots in remote places on the solar disk at the same time. He referred to \citeauthor{La_Hire_Memoires_1730}'s hypothesis of solid objects emerging in a liquid matter and explained that a scatter in the rotation period determined for different sunspots resulted from the solid objects being different in mass and showing one side or the other.

\citet{Cassini_1700} wrote: ``\textit{If you want the spots to be new generations} [i.e. not recurrent], \textit{this is another idea; it will resemble rather an impure drink that boils and throws up foam, than a sea in which a foreign body swims. But how is it possible to find in this system} [foam hypothesis] \textit{so many regular returns of the same spot after so many years?}". Thus, Giovanni Domenico Cassini was inclined to the idea of recurrent sunspot emergence. 

Next in the chronology is the article by \citet{La_Hire_1701}, who followed his hypothesis of recurrent solids and calculated that the sunspots in November and December 1700 were separated by 54 days and therefore rotated with a period of about 27 days, i.e. close to his own estimates five years ago.

\citet{Cassini_Maraldi_1701} wrote that 142 days elapsed between 7 November 1700 when the sunspot passed through the middle of the Sun (by calculation, not actual observation) and 29 March 1701, which gives 5 rotations of 28.4 days (``\textit{28 jours 2/5}"), if it was the same sunspot.

Jacques \citet{Cassini_1701}, a son of Giovanni Domenico Cassini, wrote that the observations in November 1700, March 1701, and November 1701 gave him the occasion to investigate whether the apparent motion deduced from these observations fulfilled the new hypothesis rejecting his father's hypothesis of the motion of sunspots (i.e. a theory of the solar precession). He described his improvements of the precession of the Sun relative to the ecliptic and other constellations and reconstructed the sunspot trajectory (Figure~\ref{Fig10}e) with a smaller scatter in latitude than they had previously obtained, as he stated. He then recalculated the time of the sunspot passage through the central meridian on 29 March and 5 November 1701 and concluded that the sunspot made 8 rotations with a period of 27 days and 14.5 hours, assuming it was the same sunspot. 

Finally, we would like to refer to the German observations. \citet{2018AN....339..219N} translated a note from Wurzelbaur's letter about some faint spots on 10 to 13 June 1700. \citeauthor{La_Hire_1697_1704} measured the altitude and the limb passages on 10, 12, and 13 June, but did not give any notes on sunspots. Presumably, small spots were not resolved by a small telescope used for the altitude measurements. A higher magnification telescope was not routinely used, since the observers did not expect recurrent sunspots on those days.

\citet{2018AN....339..219N} also reconstructed the sunspot latitude of $3.2 \pm 7.2^{\circ}$ from 7 to 13 November 1700 based on another Wurzelbaur's letter. We compare the text description of a sunspot group in November given by Wurzelbaur with the engravings by \citet{La_Hire_1700} and conclude that they reported the same active region.

Kirch noted two spots on 31 December 1700. We now understand that it has been a complex active region. In his letter to Kirch, Ihle reported one sunspot on 3, 7, and 9 November 1701, but a spotless Sun in October and on 10 November. Figure~\ref{Fig10}e incorporates the sunspot track reconstructed by \citet{2018AN....339..219N} in yellow. They evaluated the sunspot latitude to be $-12.4 \pm 8^{\circ}$. On 3 November 1701, the sunspot position was given by Ihle imprecisely and can be improved.

It is valuable that Ihle also indicated that the sunspot was nicely round with size of 1/60 of the solar diameter. We scale the yellow circles in Figure~\ref{Fig10}e to match this value. If Ihle defined a sunspot by its umbra as many astronomers did \citep{2025SoPh..300...17Z}, the sunspot drawn in \citeauthor{La_Hire_1697_1704}'s journal has a more correct size than the one engraved in \citeauthor{Cassini_1701} (Figure~\ref{Fig10}c and e).

\subsection{May\,--\,December 1702}
\label{Tome7_2}

\begin{figure}    
\centerline{\includegraphics[width=1\textwidth,clip=]{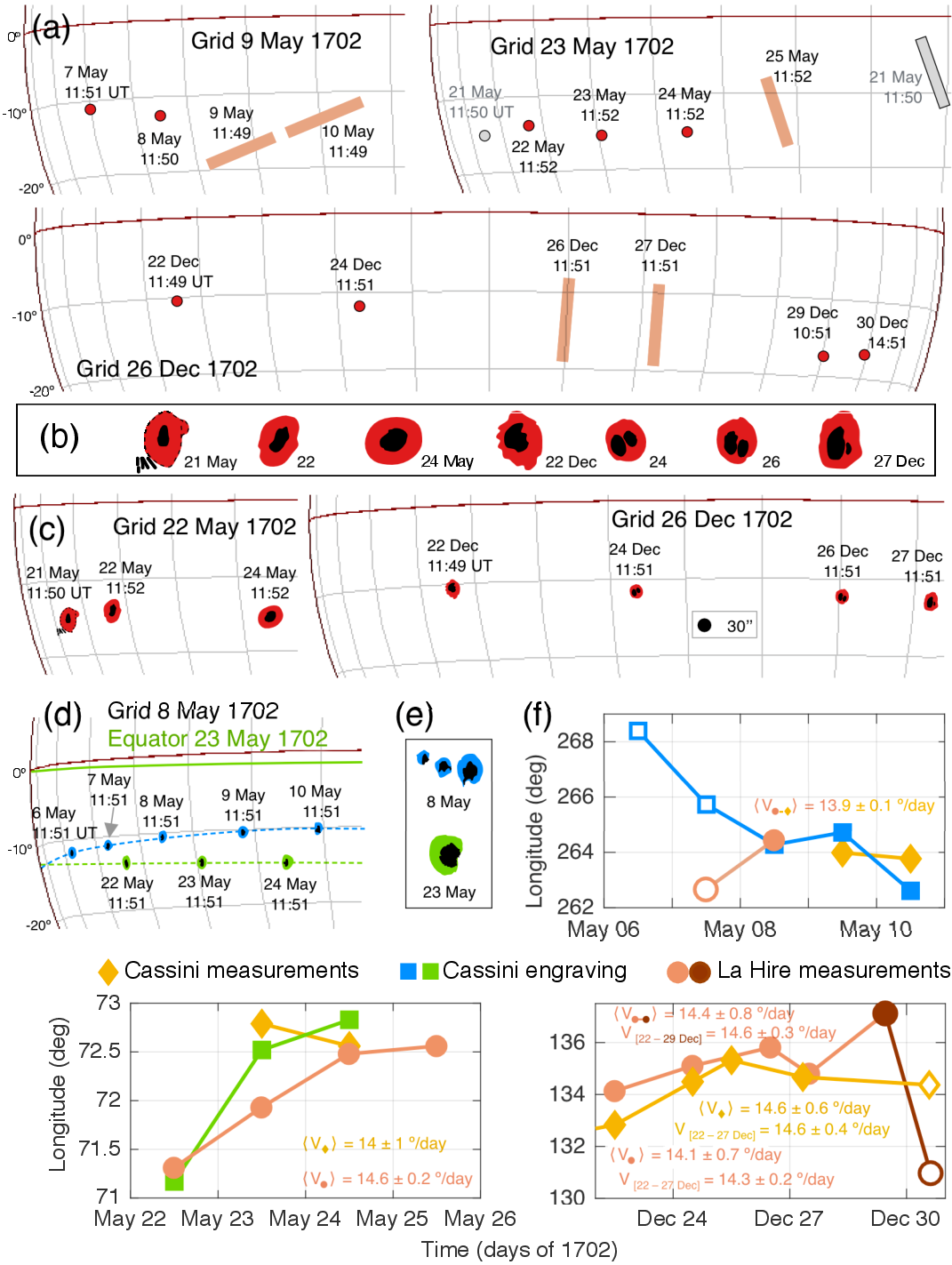}}
\small
        \caption{(\textbf{a}) Sunspot positions derived from the measurements by \citet{La_Hire_1697_1704}. (\textbf{b}) Horizontally and vertically mirrored sunspot reproductions drawn by \citeauthor{La_Hire_1697_1704}. (\textbf{c}) Reconstructed sunspot transits. The black circle is a $30''$ object. (\textbf{d}) Color-enhanced modification of the original engraving from \citet{Cassini_1702} rotated by $7^{\circ}$. The heliographic grid corresponds to 8 May and the Equator position to 23 May 1702. (\textbf{e}) Horizontally flipped sunspot engravings from \citet{Cassini_1702}. (\textbf{f}) Longitudes and rotation rate reconstructed from the measurements and engraving by \citeauthor{La_Hire_1697_1704} and Cassini. \textit{Empty symbols} denote unreliable data.}
\label{Fig11}
\end{figure}

Figure~\ref{Fig11}a represents sunspot positions, reconstructed from \citeauthor{La_Hire_1697_1704}'s measurements. On 11 May 1702, the sunspot was observed, but \citeauthor{La_Hire_1697_1704} could not measure it. On 21 May, the sunspot passage was registered with an error, because the sunspot appeared close to the disk centre (gray rectangle). The gray circle shows the presumed location of the sunspot on that day. 

Due to bad weather, the observations stopped on 15 December until 22 December, so \citeauthor{La_Hire_1697_1704} had not seen the first appearance of the sunspot. On 29 and 30 December, the sunspot was located at higher latitude which was mentioned in \citet{La_Hire_1703}.

On the same dates, \citet{Cassini_1703} calculated the sunspot latitude of $-10$ or $-11^{\circ}$ in the first observations in December and further noted that there were some irregularities in the sunspot position at the end of December (see Appendix~\ref{Tome7_1_app}). We take a quick look at \citeauthor{Cassini_1702_1704}'s observational journal and find that the latitude of the sunspot was measured only on 22 and 24 December, and he probably referred to \citeauthor{La_Hire_1697_1704}'s measurements.

Thus, there was either one sunspot group from 22 to 30 December and the difference in latitude is due to poor weather conditions, or there were two sunspot groups at different latitudes. Note that \citeauthor{La_Hire_1697_1704}'s approach to micrometer measurements on 29 and 30 December changed as compared to his routine measurements on other days, which may explain possible uncertainties in sunspot positions.

In Figure~\ref{Fig11}a, we get the average latitude of $-11.9 \pm 0.7^{\circ}$ on 7 and 8 May, $-14.9 \pm 0.6^{\circ}$ on 22\,--\,24 May, $-10.9 \pm 0.7^{\circ}$ on 22 and 24 December, and $-17.4 \pm 0.3^{\circ}$ for 29 and 30 December 1702. 

\citet{Spoerer1889} extrapolated 1\,--\,2 day measurements from 
Mémoires and listed sunspot latitudes as $-10.5^{\circ}$ from 6 to 11 May, $-12^{\circ}$ from 20 to 25 May, and $-11^{\circ}$ from 22 to 31 December 1702.

Figure~\ref{Fig11}b illustrates horizontally and vertically mirrored sunspot reproductions from \citeauthor{La_Hire_1697_1704}'s journal. Sunspots were drawn schematically with ink. On 21 May 1702, the dashes behind the sunspot could indicate either trailing spots or faculae. On 22 May, next to the sunspot drawing, there is a note saying that it was observed with a 16-foot telescope (\textit{visa per tubum 16 ped}). There is a notable difference in the umbra orientations on 24 and 26 December, but overall similarities in the fine structure of the sunspot from 22 to 27 December suggests that this active region is most likely the same object.

Figure~\ref{Fig11}c shows the reconstructed sunspot tracks based on \citeauthor{La_Hire_1697_1704}'s drawings and measurements. Sunspots were not drawn by \citeauthor{La_Hire_1697_1704} from 7 to 10 May. From 21 to 24 May, the sunspots were sketched without any hint to their size relative to the solar disk. Here, we rely on the size of this sunspot in \citeauthor{Cassini_1702_1704}'s journal where the sunspot is sketched with the limb. Recall however that sketches often overestimate the size of sunspots.

In his journal, \citeauthor{La_Hire_1697_1704} had not provided any information on the sunspot size in December. Figure~\ref{Fig11}c contains a $30''$-object to which we scale the horizontal size of the sunspot penumbra in December 1702 following the notes in Mémoires \citep{La_Hire_1703}.

In Mémoires, \citet{La_Hire_1703} noted that the sunspot was of ``mediocre" size and composed of two principal spots which were surrounded by a kind of cloud. The diameter of the whole mass around the spot was about half of an arcminute, and the largest of the two spots was always observed, as we translate and retell from French. We would like to hypothesize that either La Hire had a different notebook, or the sunspot size information was provided by another astronomer at the observatory, because, on those days, his journal has no notes on the sunspot size.

Figure~\ref{Fig11}d illustrates part of the color-enhanced modification of the original engraving of the whole solar disk from \citet{Cassini_1702} rotated by $7^{\circ}$ and with the heliographic grid of 8 May and the Equator position from 23 May 1702 imposed. This engraving, showing only the leading sunspot transits, was accompanied by the individual sunspot engravings on 8 and 23 May. Their horizontally mirrored reproductions are shown in Figure~\ref{Fig11}e. We preliminary check how the sunspot group on 8 May was drawn in Cassini's hand-written journal and revealed that its orientation is almost the same as it was engraved in Mémoires. Thus, we may conclude that this active region either had a zero tilt angle or slightly violated Joy's law. The size of the sunspots in Figure~\ref{Fig11}d is slightly less than half of an arcminute.

Figure~\ref{Fig11}f depicts the longitudes. Empty symbols represent less reliable values derived for near-limb sunspots. The blue and green squares mark the longitudes reconstructed from the engraving in Figure~\ref{Fig11}d. We do not estimate the rotation rate from them since engravings are commonly less reliable than measurements.

The reddish circles correspond to \citeauthor{La_Hire_1697_1704}'s measurements. Yellow diamonds are the sunspot longitudes estimated by \citet{Cassini_1702_2, Cassini_1702, Cassini_1703} and rounded by him to $0.5^{\circ}$. Combining one measurement by \citeauthor{La_Hire_1697_1704} on 8 May and two measurements by \citeauthor{Cassini_1702} on 9 and 10 May, we get $V_{[8-10May]} = \left\langle V_{\bullet {\scriptscriptstyle \blacklozenge}} \right\rangle = 13.9 \pm 0.1^{\circ}$~d$^{-1}$.  On 23\,--\,24 May, \citeauthor{Cassini_1702}'s measurements yield $\left\langle V_{\scriptscriptstyle \blacklozenge} \right\rangle = 14.0 \pm 1.0^{\circ}$~d$^{-1}$. On 22 to 25 May, \citeauthor{La_Hire_1697_1704}'s measurements give $V_{[22-25May]} = \left\langle V_{\bullet} \right\rangle = 14.6 \pm 0.2^{\circ}$~d$^{-1}$.

In December, \citeauthor{Cassini_1702}'s measurements yield $\left\langle V_{\scriptscriptstyle \blacklozenge} \right\rangle = 14.6 \pm 0.6^{\circ}$~d$^{-1}$ and $V_{[22-27Dec]} = 14.6 \pm 0.4^{\circ}$~d$^{-1}$. For \citeauthor{La_Hire_1697_1704}, we obtain different estimates depending on whether we assume that one or two sunspot groups were observed. If the sunspot on 22\,--\,27 December is the same as on 29\,--\,30 December, and the jump in latitude was caused by bad weather, then $\left\langle V_{\bullet} \right\rangle = 14.4 \pm 0.8^{\circ}$~d$^{-1}$ and $V_{[22-29Dec]} = 14.6 \pm 0.3^{\circ}$~d$^{-1}$. If there were two sunspot groups, then $\left\langle V_{\bullet} \right\rangle = 14.1 \pm 0.7^{\circ}$~d$^{-1}$ and $V_{[22-27Dec]} = 14.3 \pm 0.2^{\circ}$~d$^{-1}$.

The rotation rate between the sunspots on 25 May and 22 December 1702 is $V_{[25May-22Dec]} = 14.5 \pm 0.1^{\circ}$~d$^{-1}$. However, it is difficult to combine them into a nest of activity, because of the large time gap.

We now summarize the relevance of the observations and measurements published by La Hire and Cassini's son in Mémoires. \citet{Cassini_1702} wrote that they observed the sunspot from 6 to 11 May 1702. On 7 May, the spot was seen using a 45-foot telescope to consist of two spots joined together, of which the smaller was near the eastern edge. It was surrounded by penumbra and faculae. On 8 May, it consisted of three sunspots separated from each other. On 9 and 10 May, there was no noticeable change in their configuration. The text is as follows: ``\textit{On the morning of 11 May the spot was about the same size in a 17-foot telescope, but much less dark, so it was difficult to distinguish. At 10 o'clock, after trying to determine its position, I could not see it in a 6-foot telescope, which is by the parallactic machine, and in the evening it could not be distinguished in the 45-foot telescope. What is unusual about this observation is that it was the decrease in its visibility, not its size, that caused it to disappear.}"

\citeauthor{Cassini_1702} provided information about the sunspot trajectory across the solar disk, calculated the sunspot central meridian distance and concluded that the obtained values were consistent with the daily motion of the spot, which was just over $13^{\circ}$~d$^{-1}$ in synodal units\footnote{If we take the synodal rotation period of the heliospheric grid to be 27.2753 days, the synodal rotation rate is $13.2^{\circ}$~d$^{-1}$.}.

\citeauthor{Cassini_1702} turned to the observations in November 1700, March and November 1701. He calculated that 550.5 days passed between the sunspot crossing of the central meridian on 7 November 1700 and 11 May 1702, giving a period of 27 days 12 hours and 35 minutes. He then turned to the sunspots in May 1695, May 1684, 1686 and 1688 and calculated that between 6 May 1688 and 11 May 1702 there were 182 rotations lasting 27 days 12 hours and 21 minutes. 

\citeauthor{Cassini_1702} also referred to his father's calculations, according to which there were 783 revolutions between the sunspot of 19 May 1625 observed by Scheiner and the sunspot in May 1684, which is 27 days 12 hours and 20 minutes. He also combined the sunspots observed by Hevelius in May 1644 and 13 May 1688 and sunspots of 1686 and 1688. Thus, he counted six large time intervals with very similar rotation periods. Here, it is worth pointing out that many sunspots were recorded in May at nearly $-10^{\circ}$ latitudes from 1684 to 1702. We assume that the Parisian observers expected sunspots to appear in May and used large telescopes more often, otherwise they commonly used a small telescope.

The article by \citet{La_Hire_1702} said that at the beginning of May 1702 a small spot (\textit{une petite tache}) appeared. It disappeared after a few days, gradually decreasing. Recall that \citeauthor{Cassini_1702} gave another description. \citeauthor{La_Hire_1702} continued that on 21 May another single spot, about the same size as the previous one, appeared on the eastern limb and on 25 May it was so faint that it could hardly be seen in a 3-foot telescope.

\citet{La_Hire_1702} pointed out that sunspots spaced far apart on the solar disk are very rarely seen, and the sunspots in the beginning and in the end of May 1702 are not the same. Otherwise, the rotation period would be 14 days. \citeauthor{La_Hire_1702} tried to explain the period obtained in the framework of his hypothesis of one or more solids surfacing in liquid matter. Recall that Cassini stated that he never saw more than one sunspot on the solar disk, apparently meaning that he never saw two distant sunspots.

\citet{Cassini_1702_2} reported that on 22 May 1702 a new spot appeared, it was larger and darker (\textit{obscure}) than the previous one. This differs from \citeauthor{La_Hire_1702}'s description. \citeauthor{Cassini_1702_2} then wrote that the sunspot had a penumbra. The next day, there was a single sunspot as seen through a 17-foot telescope. On 24 May, it appeared to be composed of two joined sunspots and was a bit smaller. On 25 May, the sunspot was difficult to distinguish, because it was very faint and not very obscure. It could not be seen in a 6-foot telescope at noon, and was observed since. Further, \citeauthor{Cassini_1702_2} described the sunspot trajectory and gave its coordinates. He calculated that the spot should have already been on the disk on 20 May, but he could not see it, no matter how hard he tried. Thus, apparently, Cassini made an extra effort to observe spots in May each year.

Since the longitude of the spot on 22\,--\,24 May 1702 differed a lot from the four previous spots observed two years ago, \citeauthor{Cassini_1702_2} combined it with the sunspot reported by Maraldi in May 1695 and got the period of 27 days 12 hours and 2 minutes. He concluded that either it was the same sunspot or at least the sunspots formed in the same place. 

\citet{Cassini_1703} listed the coordinates of the sunspot from 22 to 30 December 1702. On 30 December, the sunspot was specified as still appearing very large, although it was quite close to the limb. \citeauthor{Cassini_1703} predicted that the sunspot may reappear on 15 January 1703. 

On 22 December, \citet{La_Hire_1703} described the sunspot to be of ``mediocre" size and consisted of two principal spots with penumbra. He listed the sunspot coordinates and described the sunspot trajectory.

Finally, we would like to mention German observations from 22 to 28 December \citep{2018AN....339..219N}. The sunspot positions given in inches or digits by Ihle and Wurzelbaur allow us to reconstruct the longitudes. The raw estimates agree well with \citeauthor{La_Hire_1697_1704}'s measurements and suggest that the German and French astronomers reported the same active region.

\section{Discussion}
\label{Discussion}

\begin{figure}    
\centerline{\includegraphics[width=1\textwidth,clip=]{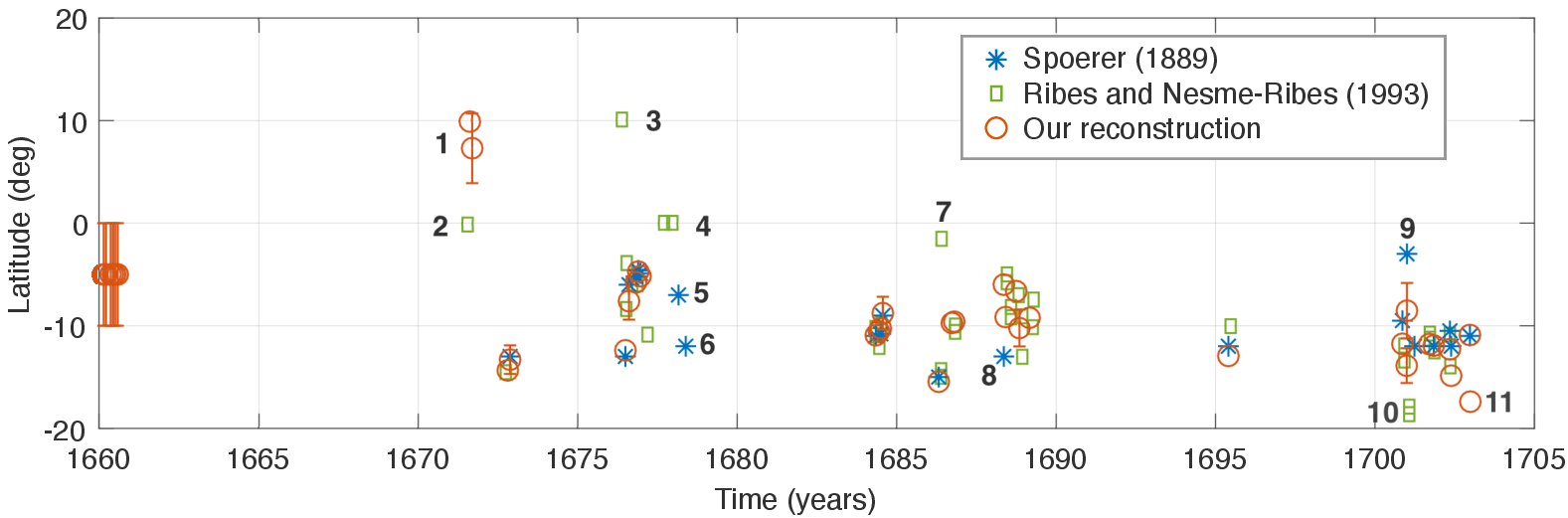}}
\small
        \caption{Time-latitude distribution of sunspots in 1660\,--\,1702: \citet{Spoerer1889} is shown by blue asterisks, \citet{1993A&A...276..549R} by green rectangles, and our reconstruction by red circles. Numbers indicate notable discrepancies between the reconstructions.}
\label{Fig12}
\end{figure}

Figure~\ref{Fig12} compares the time-latitude distributions of sunspots according to \citet{Spoerer1889} by blue asterisks, \citet{1993A&A...276..549R} by green rectangles, and our results by red circles. \citeauthor{Spoerer1889} provided the latitudes of the largest-area sunspots in a group. For our reconstruction, if an observer had measured the position of several sunspots in a group, we plot the average latitude of the sunspot group. We also show results for 1660 obtained in \citet{2025SoPh..300...17Z}.

If a sunspot was observed for several days, we calculate its average latitude and place it in the middle of this time interval. Error bars are only shown where the uncertainty in sunspot latitude is large. When a sunspot was observed by several observers, we show the data obtained only from the Parisian observations. Thus, each series of sunspot observations is shown by one point. Apparently, \citeauthor{1993A&A...276..549R} also plotted the average latitude. To recover sunspot latitudes from  \citet{1993A&A...276..549R}, we correct Figure~6 therein with an image-editing software and extract sunspot positions.

We would like to comment on notable differences between the reconstructions:
\begin{enumerate}
\item These observations were made by Cassini from 11 to 19 August 1671 and by Heinrich Siverus from 5 to 15 September 1671 and share similar longitudes. The latter observations are more uncertain, and so the sunspot latitude may be about $10^{\circ}$ in both cases.
\item Apparently, this point came from Cassini's observations in August\,--\,September 1671. It is unclear why \citeauthor{1993A&A...276..549R} placed sunspots almost at the Equator.
\item This point corresponds to the observations in 26 June\,--\,4 July 1676. The cause of the sunspot in the northern hemisphere is unclear to us.
\item In all likelihood, these two rectangles came from Cassini's observations in 1677 and 1678, which we have not analyzed yet.
\item These observations were done in 25 February\,--\,4 March 1678. Discrepancy with Nesme-Ribes's reconstruction is prominent. We have not analyzed yet Cassini's journals and the corresponding publication in Mémoires.
\item The same as in the previous comment. The point is based on observations made from 21 to 30 May 1678.
\item The observation in 1686. The reason why the spot appeared close to the Equator is not clear yet.
\item The observation of 12 May 1688, \citeauthor{Spoerer1889} gave a latitude of $-13^{\circ}$. This estimate is based on the following text by \citet{Cassini_Memoires_1730_1688}: ``\textit{On that day at six o'clock in the morning, having observed the Sun through a 3-foot telescope... he saw two spots in the western part of the disk of the Sun, of which the one closest to the edge was a little more than a sixth of its} [disk] \textit{diameter away. It was also at the same distance from the diameter of the Sun parallel to the Equinox. The other spot was carried away by the universal movement to the West, on the same line parallel to the Equinox, in such a way that the distance between these two spots was only one minute or a little more.}" as we translate from French. These measurements are consistent with those by \citet{La_Hire_1687_1689}, if the distance is taken from a horizontal line passing through the centre of the disk. We suggest that \citeauthor{Spoerer1889} used the $7^{\circ}$-inclined line, therefore, he got more southward latitude compared to our results.
\item On 2 January 1701, \citeauthor{Spoerer1889} indicated a latitude of $-3^{\circ}$. This estimate comes from the following text by \citet{La_Hire_1701}: ``...\textit{on 2 January 1701, they} [sunspots] \textit{were very close to the parallel which passed through its} [disk] \textit{centre.}" Precise measurements by \citet{La_Hire_1697_1704} obtained for the trailing spot gave $-5.28^{\circ}$. The sunspot group itself was broad in latitude (Figure~\ref{Fig10}c). We estimate the sunspot-group latitude to be $-8.53^{\circ}$.
\item These points apparently correspond to the sunspot observed in 29\,--\,30 March 1701. We have not analyzed these observations since La Hire did not observe this sunspot. However, we roughly estimate the latitude to be $-12^{\circ}$ based on the engraving in \citet{Cassini_1701}. This value agrees with that in \citet{Spoerer1889}.
\item On 29 and 30 December 1702, we obtain $-17.4^{\circ}$ using the measurements by \citet{La_Hire_1697_1704}. The observations were carried out under poor weather conditions (Section~\ref{Tome7_2}).
\end{enumerate}

Table~\ref{Tab1} lists the periods when the sunspots were observed. We highlight in green those intervals when a sunspot group obeyed Joy's law and in red when it violated the law. Potentially the tilt-angles were also in agreement with Joy's law on 11\,--\,19 Aug 1671, 18\,--\,26 Oct 1672, and 28 June\,--\,9 July 1684, and violated it on 7\,--\,11 May 1702. However, the reliability of the corresponding data was too low. Recall that even if the trailing sunspots did emerge they were not always noticeable. 

\begin{table}
\caption{List of sunspot observations from 1671 to 1702: Sunspot groups obeying Joy's law are shown in green and those violating the law are shown in red.}
\label{Tab1}
\begin{tabular}{ccc}     
\toprule          

11\,--\,19 Aug 1671 & \cellcolor[HTML]{C2FCAC} 11\,--\,13 Jun 1684 & 1\,--\,5 Mar 1689 \\
\hline
5\,--\, 15 Sept 1671 & 28 Jun\,--\,9 Jul 1684 &\cellcolor[HTML]{FBC1C8} 27\,--\, 30 May 1695 \\
\hline
18\,--\,26 Oct 1672 & 25\,--\, 28 Jul 1684 & \cellcolor[HTML]{C2FCAC} 9\,--\,13 Nov 1700 \\
\hline
12\,--\,22 Nov 1672 &\cellcolor[HTML]{C2FCAC} 23 Apr\,--\,1 May 1686 &\cellcolor[HTML]{FBC1C8} 30 Dec\,--\,2 Jan 1701 \\
\hline
\cellcolor[HTML]{FBC1C8} 26 Jun\,--\,4 Jul 1676 &\cellcolor[HTML]{FBC1C8} 24\,--\,26 Sept 1686 & 15\,--\,16 Sept 1701 \\
\hline
4\,--\,15 Aug 1676 & 22 Oct 1686 & \cellcolor[HTML]{C2FCAC} 1\,--\,10 Nov 1701 \\
\hline
27 Oct\,--\,2 Nov 1676 &\cellcolor[HTML]{C2FCAC} 12 May 1688 & 7\,--\,11 May 1702 \\
\hline
19\,--\,30 Nov 1676 &\cellcolor[HTML]{C2FCAC} 1\,--\,3 Jun 1688 & 21\,--\,25 May 1702 \\
\hline
15\,--\,27 Dec 1676 & 1\,--\,2 Oct 1688 & 22\,--\,30 Dec 1702 \\
\hline
\cellcolor[HTML]{C2FCAC} 5\,--9 May 1684 & \cellcolor[HTML]{FBC1C8} 7\,--\,8 Nov 1688 &  \\
\toprule          
\end{tabular}
\end{table}
  
\begin{figure}    
\centerline{\includegraphics[width=0.5\textwidth,clip=]{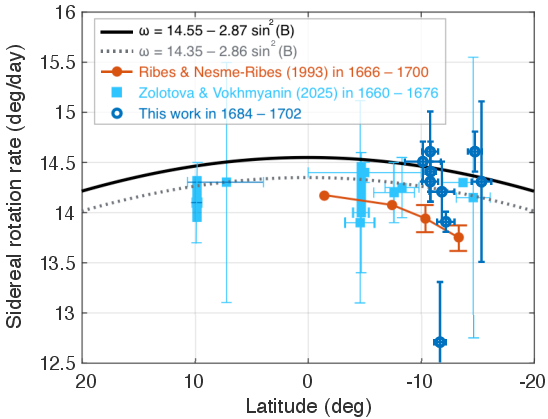}}
\small
        \caption{Rotation rate from \citet{1993A&A...276..549R} in 1666\,--\,1700, shown in \textit{red}, those from \citet{2025SoPh..300...17Z}, shown in \textit{cyan}, and those derived in this study, shown in \textit{blue}. The \textit{black solid curve} represents the rotation law derived by \citet{1986A&A...155...87B} for the full range of sunspot groups. The \textit{gray dotted} curve represents the solar rotation for large long-lived sunspot groups as derived by \citet{2018AstL...44..202N}.}
\label{Fig14}
\end{figure}

Figure~\ref{Fig14} compares the sidereal rotation rates. The results from \citet{1993A&A...276..549R} for the period of 1666\,--\,1700 are presented in red. \citet{1988IAUS..123..227R} concluded the rotation was slower, and the rotation profile was more differential. The rotation rates obtained in \citet{2025SoPh..300...17Z} for the period of 1660\,--\,1676 are shown in cyan, and those estimated in this study in blue. In the previous sections, we use two approaches to estimate the rotation rate: the average value $\left\langle V_{\bullet} \right\rangle$, which is the sum of the rotation rates between adjacent observations divided by their total number, and the rotation rate between distant measurements $V_{[{Start \hspace{1mm} time-End \hspace{1mm}time}]}$ which is less prone to uncertainties and is the true mean rotation rate in a given time interval. Both approaches give rather close results, but we consider the second one to be more correct. 

The estimates in Figure~\ref{Fig14} are based only on subjectively reliable data, i.e., those derived from three or more sunspot position measurements. Remarkably, sunspots observed in 1684\,--\,1702 occupy a very narrow latitude range. These are the following estimates:
\begin{itemize}
\item From Flamsteed's observations from 5 to 13 May 1684, we calculate the rotation rate to be $V_{[7-13May]} = 14.4 \pm 0.3^{\circ}$ d$^{-1}$ at $-9.8 \pm 0.7^{\circ}$ latitude. Flamsteed also reported sunspots on 1 June and 4 July 1684.

\item From La Hire's measurements from 5 May to 28 July 1684, we evaluate the rotation rate of $V_{[7-9May]} = 14.1 \pm 1^{\circ}$ d$^{-1}$ at $-10.9 \pm 0.5^{\circ}$ latitude, $V_{[11-12Jun]} = 14.0 \pm 1^{\circ}$ d$^{-1}$ at $-10.6 \pm 0.8^{\circ}$ (this estimate for the sunspot group), $V_{[29Jun-7Jul]} = 14.5 \pm 0.2^{\circ}$ d$^{-1}$ at $-10.3 \pm 1.6^{\circ}$, and $V_{[26-28Jul]} = 14.2 \pm 1^{\circ}$ d$^{-1}$ at $-8.8 \pm 1.6^{\circ}$.

\item From La Hire's measurements from 23 April to 22 October 1686, the rotation rate is $V_{[24-28Apr]} = 14.3 \pm 0.8^{\circ}$ d$^{-1}$ at $-15.5 \pm 0.9^{\circ}$, $V_{[24-26Sept]} = 14.0 \pm 1.0^{\circ}$ d$^{-1}$ at $-9.7 \pm 0.4^{\circ}$. Sunspot on 22 October 1686 was located at $-9.6^{\circ}$. Measurements published in Mémoires \citep{La_Hire_Memoires_1730} yield a rotation rate of $V_{[22-28Apr]} = 14.1 \pm 0.4^{\circ}$ d$^{-1}$ at $-13.6 \pm 0.8^{\circ}$.

\item From La Hire's measurements from 12 May 1688 to 5 March 1689, the sunspot group latitude on 12 May is $-6.0 \pm 1.2^{\circ}$; the rotation rate of the leading sunspot is $V_{[1-2Jun]} = 13.9 \pm 1.0^{\circ}$ d$^{-1}$ at $-8.9 \pm 0.3^{\circ}$ and of the trailing sunspot is $V_{[2-3Jun]} = 14.2 \pm 1.0^{\circ}$ d$^{-1}$ at $-9.4 \pm 0.2^{\circ}$; for the sunspot in March $V_{[1-4March]} = 14.4 \pm 1^{\circ}$ d$^{-1}$ at $-9.2 \pm 0.1^{\circ}$.

\item From 27 to 30 May 1695, based on La Hire's measurements, the rotation rate at $-12.9 \pm 1^{\circ}$ latitude is $V_{[27-28May]} = 14.2 \pm 1.0^{\circ}$ d$^{-1}$.

\item From La Hire's measurements from 9 November 1700 to 10 November 1701, the leading sunspot rotation rate is $V_{[9-12Nov \hspace{1mm} 1700]} = 12.7 \pm 0.6^{\circ}$ d$^{-1}$ at $-11.8 \pm 0.5^{\circ}$ and $V_{[1-9Nov \hspace{1mm} 1701]} = 14.2 \pm 0.3^{\circ}$ d$^{-1}$ at $-11.9 \pm 1.1^{\circ}$.

\item From La Hire's measurements from 9 May to 30 December 1702, the rotation rate is $V_{[22-25May]} = 14.6 \pm 0.2^{\circ}$ d$^{-1}$ at $-14.9 \pm 0.6^{\circ}$ and $V_{[22-27Dec]} = 14.3 \pm 0.2^{\circ}$ d$^{-1}$ at $-10.9 \pm 0.7^{\circ}$. Measurements by \citet{Cassini_1702, Cassini_1702_2, Cassini_1703} rounded by him to $0.5^{\circ}$ yield $V_{[23-24May]} = 14.0 \pm 1^{\circ}$ d$^{-1}$ and $V_{[22-27Dec]} = 14.6 \pm 0.4^{\circ}$ d$^{-1}$. Combining measurements by La Hire and Cassini, we calculate the rotation rate to be $V_{[8-10May]} = 13.9 \pm 0.1^{\circ}$ d$^{-1}$ at $-12.3 \pm 0.8^{\circ}$.
\end{itemize}

The modern rotation profile, for the full range of Greenwich sunspot groups, is represented by the solid black curve \citep{1986A&A...155...87B}. The rotation profile derived from regular long-lived sunspot groups in \citet{2018AstL...44..202N} is shown with the gray dotted curve. The two modern profiles represent the rotation of the sunspot-group, while historical rotation rates are obtained for the largest sunspot in a group. \citet{1984ApJ...283..373H} found that sunspot groups rotate slower than individual spots by about 1\%. With this correction the sunspot rotation during the Maunder minimum is in agreement with the rotation rate of regular long-lived sunspots observed in the modern era.

The rotation rate obtained in this work differs substantially from that obtained by \citet{1993A&A...276..549R} for the sunspots above $-10^{\circ}$. Currently, we do not see any clear reason for such a discrepancy. If we also consider unreliable data, we will still get significantly faster rotation rate of $14.1 \pm 0.4^{\circ}$ d$^{-1}$ at $-11.0 \pm 2.7^{\circ}$ latitude as compared to \citet{1993A&A...276..549R}. 

\section{Conclusions}
\label{S-Conclusions}

In this study, we analyze sunspot observations made between 1684 and 1702. Most of the measurements are from La Hire's journals. We carefully investigate all text notes and drawings to evaluate their reliability and to accurately restore the coordinates and areas of sunspots and their groups. We also estimate the sidereal rotation rate of sunspots and find that it is consistent with the rotation rate of regular sunspots in the modern era.

We find that the bulk of sunspot observations published by Flamsteed, La Hire, and father and son Cassini, were carried out to measure the period of the Sun's rotation and to improve the theory of the precession of the Sun's rotation axis. Both Cassini and La Hire believed that all sunspots are recurrent over years and decades. La Hire hypothesized that sunspots are caused by solids emerging in the liquid solar matter. The sunspot position measurements were taken by La Hire by means of a 3-foot telescope, which was mainly exploited for the solar altitude measurements. The fine structure of sunspots was drawn with pencil using a larger telescope, usually during morning and evening hours. The raw schematic drawings appear to have been made with a small telescope.

Comparing the sunspot longitudes obtained from the French observations in this study with the reconstructed measurements by German astronomers \citep{2018AN....339..219N}, we find that German and French observers reported the same active regions. The use of a small telescope caused La Hire to miss several small sunspots reported by German astronomers. This also means that notes about the absence of sunspots over a long period of time should be treated with caution. 

Comparison of the temporal distribution of the average sunspot latitudes, reconstructed in the previous studies, reveal a good agreement with \citet{Spoerer1889}, but several discrepancies with \citet{1993A&A...276..549R}.

We also find that the sunspot groups during the analyzed period obey Joy's law more often than they violate it.

For the benefit of open discussion, all processed drawings are available at \url{http://geo.phys.spbu.ru/~ned/History.html}.


 \appendix   

\section{Flamsteed 5 May\,--\,8 July 1684}
\label{Flams_app}

To determine sunspot coordinates, for all observations except the first one, we set the disk diameter of $31'$ $28''$. The observations lasted 10 to 30 minutes, introducing a sunspot mapping uncertainty that increases toward the limb \citep[for details]{2025SoPh..300...17Z}. Flamsteed counted the hours of the day from noon. On the first two days, there were presumably date errors. The deduced longitudes would be inconsistent with the subsequent observations based on the noted date. We correct the date and added two actual observations on 5 and 6 May from \citet{1684RSPT...14..535F}.

On 9 June 1684, Flamsteed noted the distance to an earlier sunspot (\textit{Macula antecedens}), which can be interpreted as a leading spot or a sunspot from the previous rotation. He wrote that the observation was very difficult because of clouds, wind, and the high altitude of the Sun. 

\section{La Hire, Journal from 25 January 1683 to 10 October 1684}
\label{Tome1_app}

The solar diameter must be known to determine sunspot coordinates from angular distance measurements. We interpolate the routine measurements from \citet{Monnier_1741} in 1666\,--\,1670 and determine the angular size of the solar disk was $31'$ $48''$ on 5\,--\,9 May and $31'$ $40''$ on 11\,--\,13 June. We use $31'$ $42''$ from 28 June to 28 July, based on the diameter measurement in \citeauthor{La_Hire_1683_1684}'s journal on 28 June. We estimate the accuracy of the solar diameter measurement to be about $5''$ for this period, which introduces a $0.5^{\circ}$-uncertainty to the heliographic grid with the strongest impact on sunspot mapping near the limb.

Based on the midday solar passages, we assume that on 5 May 1684 the sunspot should actually be $13'$ $52''$ higher than it was noted, as otherwise the reconstructed sunspot location is outside the solar disk.

\citeauthor{La_Hire_1683_1684} did not report the time of the measurements on 8 May 1684 and gave only the angular distance between the sunspot and the limb. Matching the sunspot longitudes, we suspect that the observation was made on 9 May about 6:30 local time. Figure~\ref{Fig4}a shows the corrected sunspot position with an annular sector.

A sunspot was at the very edge, according to the measurements on 13 May 1684 (Figure~\ref{Fig4}a), though it is unclear whether it was a leading or a trailing sunspot, or there was a conjunction due to forshortening. The sunspot seems to be enlarged and distant from the limb in the schematic \citeauthor{La_Hire_1683_1684}'s drawing of the solar disk, which should be taken with caution.

On 3 and 4 July 1684, we additionally take into account that the passage was measured not with a wire (\textit{filum quadratis muralis} or \textit{mural}), but through the vertical (\textit{per verticalem}). We estimate the time lapsed between them of $14.5 \pm 0.5$~seconds\footnote{An error of one second in time measurements results in an inaccuracy of $1 – 1.5^{\circ}$ in a sunspot position at the center of the solar disk, and about $3^{\circ}$ for a spot with the central meridian distance of $60^{\circ}$.} at 3\,--\,4 hours after noon. On 4 July at noon, the preceding limb passed after the sunspot, thus we add a time gap to the vertical with the opposite sign. Figure~\ref{Fig4}a demonstrates the corresponding latitude uncertainty of 2\,--\,3$^{\circ}$.

The measurements of the solar limbs passage on 15 and 16 May 1684 contain no sunspot information. On 8 June, solar measurements also contain no sunspot information. On 27 July 1684, the passage of the solar limbs was measured, but no sunspot information was recorded.

\section{La Hire, Journal from 26 January 1686 to 30 June 1687}
\label{Tome3_app}

To determine sunspot coordinates, we interpolate the routine measurements of the Sun's size from \citet{Monnier_1741} and determine the angular size of the disk decreasing from $31'$ $54''$ on 23 April to $31'$ $50''$ on 1 May, $32'$ $9''$ in September, and $32'$ $25''$ in October 1686.

On 1 May 1686 (Figure~\ref{Fig7}a, red circle), the sunspot position was reconstructed by combining measurements with the drawing; by measurements alone, the sunspot appeared to be at the very edge of the solar disk. On 20, 21, and 22 September, and on 20, 21, 23, and 26 October, solar measurements were taken, but with no notes on a sunspot.

\section{La Hire, Journal from 1 July 1687 to 30 April 1689}
\label{Tome4_app}

We interpolate the routine measurements from \citet{Monnier_1741} and determined the angular size of the Sun's disk of $31'$ $48''$ on 12 May, $31'$ $42''$ on 1\,--\,3 June, $32'$ $11''$ on 1 and 2 October, $32'$ $32''$ on 7 and 8 November 1688, and $32'$ $21''$ on 1\,--\,5 March 1689.

On 29 and 31 May 1688, \citeauthor{La_Hire_1687_1689} took spotless measurements. On 1 June, the distance is given to the umbra (``\textit{macula(e) nigrioris}"), suggesting that umbrae may have had a penumbra. On 5, 7, and 8 June, there were spotless measurements. In the next observations, on 25 September, 3 October, 1, 3, and 6 November 1688, no spots were recorded. On 11 November, the sunspot was noticed to disappear before reaching the limb.

The sunspot position determined from the actual measurement on 4 March 1689 is illustrated with an unfilled circle in Figure~\ref{Fig8}a. Its longitude is $10^{\circ}$ bigger, compared to the measurements on 1 March. We assume that this is either an uncertainty or a typo in the measurements. If we correct the measurement from 11h-$56'$-$11"$ to 11h-$56'$-$15"$ we will obtain the position indicated by the red circle, which we use in the calculations.

\section{La Hire, Journal from 4 January 1697 to 15 August 1704}
\label{Tome7_1_app}

To determine sunspot coordinates, we use the disk angular size of $32'$ $34''$ from 9 to 13 November 1700 as measured by \citeauthor{La_Hire_1697_1704}; the same was extrapolated on 1\,--\,10 November 1701; $32'$ $44''$ from 30 December 1700 to 2 January 1701 as measured by \citeauthor{La_Hire_1697_1704}; $32'$ $2''$ on 15\,--\,16 September 1701, $31'$ $48''$ on 7\,--\,10 May 1702, $31'$ $44''$ on 21\,--\,25 May 1702, and $32'$ $44''$ on 22\,--\,30 December 1702 from interpolated measurements in \citet{Monnier_1741}.

\citet{La_Hire_1703} published his measurements in Mémoires, where he noted that the apparent motion of the sunspot in December 1702 followed a curved line (\textit{que son mouvement apparent a été en ligne courbe}), apparently referring to the higher latitudes near the western limb. 31 December was cloudy, and on 1 January 1703 the sunspot did not appear.

On the same dates in December 1702, \citet{Cassini_1703} wrote: ``\textit{In the last observations some irregularities were observed, which I do not know whether to attribute to some special movements of the spot or to the difficulty of determining its position owing to the bad weather and winds which always blew at the time, when the spot was on the visible disk of the Sun.}"

\begin{acks}

We use data from the archives of Bibliotheque l’Observatoire de Paris \url{https://bibnum.obspm.fr}, the Biodiversity Heritage Library \url{https://www.biodiversitylibrary.org}, and Bibliothèque Nationale de France \url{https://www.bnf.fr/en}.

\end{acks}

\section*{Funding}
This work has no funding.

\section*{Disclosure of Potential Conflicts of Interest}
The authors declare that they have no conflicts of interest.

\bibliographystyle{spr-mp-sola}
\bibliography{Zolotova_Vokhmyanin}  

\begin{thebibliography}{69}
\ifx\bisbn     \undefined \def\bisbn  #1{ISBN #1}\fi
\ifx\binits    \undefined \def\binits#1{#1}\fi
\ifx\bauthor   \undefined \def\bauthor#1{#1}\fi
\ifx\batitle   \undefined \def\batitle#1{#1}\fi
\ifx\bjtitle   \undefined \def\bjtitle#1{\textit{#1}}\fi
\ifx\bvolume   \undefined \def\bvolume#1{\textbf{#1}}\fi
\ifx\byear     \undefined \def\byear#1{#1}\fi
\ifx\bissue    \undefined \def\bissue#1{#1}\fi
\ifx\bfpage    \undefined \def\bfpage#1{#1}\fi
\ifx\blpage    \undefined \def\blpage #1{#1}\fi
\ifx\burl      \undefined \def\burl#1{\href{#1}{\textsf{URL}}}\fi
\ifx\href      \undefined \def\href#1#2{#2}\fi
\ifx\betal     \undefined \def\betal{et al.}\fi
\ifx\bctitle   \undefined \def\bctitle#1{#1}\fi
\ifx\beditor   \undefined \def\beditor#1{#1}\fi
\ifx\bbtitle   \undefined \def\bbtitle#1{\textit{#1}}\fi
\ifx\bedition  \undefined \def\bedition#1{#1}\fi
\ifx\bseriesno \undefined \def\bseriesno#1{\textbf{#1}}\fi
\ifx\blocation \undefined \def\blocation#1{#1}\fi
\ifx\bsertitle \undefined \def\bsertitle#1{\textit{#1}}\fi
\ifx\bsnm      \undefined \def\bsnm#1{#1}\fi
\ifx\bsuffix   \undefined \def\bsuffix#1{#1}\fi
\ifx\bparticle \undefined \def\bparticle#1{#1}\fi
\ifx\barticle  \undefined \def\barticle#1{}\fi
\ifx\binstitute  \undefined \def\binstitute#1{#1}\fi
\ifx\bpublisher  \undefined \def\bpublisher#1{#1}\fi
\ifx\doiurl    \undefined \def\doiurl#1{\href{#1}{DOI}}\fi
\makeatletter
\def\safeHref#1#2#3{\in@{http}{#2}\ifin@\href{#2}{#3}\else\href{#1#2}{#3}\fi}
\makeatother
\ifx\adsurl    \undefined
  \def\adsurl#1{\safeHref{https://ui.adsabs.harvard.edu/abs/}{#1}{ADS}}\fi
\ifx\arxivurl  \undefined
  \def\arxivurl#1{\safeHref{http://arxiv.org/abs/}{#1}{arXiv}}\fi
\ifx\botherref \undefined \def\botherref#1{}\fi
\ifx\url       \undefined \def\url#1{#1}\fi
\ifx\bchapter  \undefined \def\bchapter#1{}\fi
\ifx\bbook     \undefined \def\bbook#1{}\fi
\ifx\bcomment  \undefined \def\bcomment#1{#1}\fi
\ifx\oauthor   \undefined \def\oauthor#1{#1}\fi
\ifx\citeauthoryear \undefined\def \citeauthoryear#1{#1}\fi
\def\endbibitem {}
\ifx\bconflocation  \undefined \def\bconflocation#1{#1} \fi

\bibitem[\protect\citeauthoryear{{Académie royale des
  sciences}}{1733}]{Cassini_La_Hire_Hist_1686}
\begin{botherref}
\oauthor{\bsnm{{Académie royale des sciences}}}:
1733,
{Diverses observations Astromoniques}.
\textit{Hist. Acad. R. Sci., Depuis 1666 jusqu'a son Renouvellement 1699},
12.
\end{botherref}
\endbibitem

\bibitem[\protect\citeauthoryear{{Balthasar}, {Vazquez}, and
  {Woehl}}{1986}]{1986A&A...155...87B}
\begin{barticle}
\bauthor{\bsnm{{Balthasar}}, \binits{H.}},
\bauthor{\bsnm{{Vazquez}}, \binits{M.}},
\bauthor{\bsnm{{Woehl}}, \binits{H.}}:
\byear{1986},
\batitle{{Differential rotation of sunspot groups in the period from 1874
  through 1976 and changes of the rotation velocity within the solar cycle}}.
\bjtitle{\aap}
\bvolume{155},
\bfpage{87}.
\adsurl{1986A&A...155...87B}.
\end{barticle}
\endbibitem

\bibitem[\protect\citeauthoryear{{Bhattacharya}
  et~al.}{2024}]{2024SoPh..299...45B}
\begin{barticle}
\bauthor{\bsnm{{Bhattacharya}}, \binits{S.}},
\bauthor{\bsnm{{Lef{\`e}vre}}, \binits{L.}},
\bauthor{\bsnm{{Chatzistergos}}, \binits{T.}},
\bauthor{\bsnm{{Hayakawa}}, \binits{H.}},
\bauthor{\bsnm{{Jansen}}, \binits{M.}}:
\byear{2024},
\batitle{{Rudolf Wolf to Alfred Wolfer: the transfer of the teference observer
  in the International Sunspot Number Series (1876{\textendash}1893)}}.
\bjtitle{\solphys}
\bvolume{299},
\bfpage{45}.
\doiurl{https://doi.org/10.1007/s11207-024-02261-7}.
\adsurl{2024SoPh..299...45B}.
\end{barticle}
\endbibitem

\bibitem[\protect\citeauthoryear{{Boistel}}{2004}]{Boistel_2004}
\begin{barticle}
\bauthor{\bsnm{{Boistel}}, \binits{G.}}:
\byear{2004},
\batitle{{Nouvelle theorie des taches du Soleil, by E. Pezenas (1692-1776)}}.
\bjtitle{Cahiers François Viète, Série I}
\bvolume{8},
\bfpage{5}.
\doiurl{https://doi.org/10.4000/cahierscfv.2424}.
\burl{http://dx.doi.org/10.4000/cahierscfv.2424}.
\end{barticle}
\endbibitem

\bibitem[\protect\citeauthoryear{{Bretagnon} and
  {Francou}}{1988}]{1988A&A...202..309B}
\begin{barticle}
\bauthor{\bsnm{{Bretagnon}}, \binits{P.}},
\bauthor{\bsnm{{Francou}}, \binits{G.}}:
\byear{1988},
\batitle{{Planetary theories in rectangular and spherical variables: VSOP87
  solution.}}
\bjtitle{\aap}
\bvolume{202},
\bfpage{309}.
\adsurl{1988A&A...202..309B}.
\end{barticle}
\endbibitem

\bibitem[\protect\citeauthoryear{{Carrasco} and
  {Vaquero}}{2016}]{2016SoPh..291.2493C}
\begin{barticle}
\bauthor{\bsnm{{Carrasco}}, \binits{V.M.S.}},
\bauthor{\bsnm{{Vaquero}}, \binits{J.M.}}:
\byear{2016},
\batitle{{Sunspot observations during the Maunder minimum from the
  correspondence of John Flamsteed}}.
\bjtitle{\solphys}
\bvolume{291},
\bfpage{2493}.
\doiurl{https://doi.org/10.1007/s11207-015-0839-0}.
\adsurl{2016SoPh..291.2493C}.
\end{barticle}
\endbibitem

\bibitem[\protect\citeauthoryear{{Carrasco}, {{\'A}lvarez}, and
  {Vaquero}}{2015}]{2015SoPh..290.2719C}
\begin{barticle}
\bauthor{\bsnm{{Carrasco}}, \binits{V.M.S.}},
\bauthor{\bsnm{{{\'A}lvarez}}, \binits{J.V.}},
\bauthor{\bsnm{{Vaquero}}, \binits{J.M.}}:
\byear{2015},
\batitle{{Sunspots during the Maunder minimum from Machina Coelestis by
  Hevelius}}.
\bjtitle{\solphys}
\bvolume{290},
\bfpage{2719}.
\doiurl{https://doi.org/10.1007/s11207-015-0767-z}.
\adsurl{2015SoPh..290.2719C}.
\end{barticle}
\endbibitem

\bibitem[\protect\citeauthoryear{{Carrasco} et~al.}{2022}]{2022ApJ...927..193C}
\begin{barticle}
\bauthor{\bsnm{{Carrasco}}, \binits{V.M.S.}},
\bauthor{\bsnm{{Mu{\~n}oz-Jaramillo}}, \binits{A.}},
\bauthor{\bsnm{{Gallego}}, \binits{M.C.}},
\bauthor{\bsnm{{Vaquero}}, \binits{J.M.}}:
\byear{2022},
\batitle{{Revisiting Christoph Scheiner's sunspot records: a new perspective on
  solar activity of the early telescopic era}}.
\bjtitle{\apj}
\bvolume{927},
\bfpage{193}.
\doiurl{https://doi.org/10.3847/1538-4357/ac52ee}.
\adsurl{2022ApJ...927..193C}.
\end{barticle}
\endbibitem

\bibitem[\protect\citeauthoryear{{Carrasco} et~al.}{2024}]{2024ApJ...968...65C}
\begin{barticle}
\bauthor{\bsnm{{Carrasco}}, \binits{V.M.S.}},
\bauthor{\bsnm{{Aparicio}}, \binits{A.J.P.}},
\bauthor{\bsnm{{Chatzistergos}}, \binits{T.}},
\bauthor{\bsnm{{Jamali Jaghdani}}, \binits{S.}},
\bauthor{\bsnm{{Hayakawa}}, \binits{H.}},
\bauthor{\bsnm{{Gallego}}, \binits{M.C.}},
\bauthor{\bsnm{{Vaquero}}, \binits{J.M.}}:
\byear{2024},
\batitle{{Understanding solar activity after the Maunder minimum: sunspot
  records by Rost and Alischer}}.
\bjtitle{\apj}
\bvolume{968},
\bfpage{65}.
\doiurl{https://doi.org/10.3847/1538-4357/ad3fb9}.
\adsurl{2024ApJ...968...65C}.
\end{barticle}
\endbibitem

\bibitem[\protect\citeauthoryear{{Cassini}}{1684}]{Scavans_1684}
\begin{botherref}
\oauthor{\bsnm{{Cassini}}, \binits{G.D.}}:
1684,
{Description d'une tache qui a paru dans le Soleil ce mois de May dernier
  1684}.
\textit{Le Journal des Sçavans},
177.
\end{botherref}
\endbibitem

\bibitem[\protect\citeauthoryear{{Cassini}}{1700}]{Cassini_1700}
\begin{botherref}
\oauthor{\bsnm{{Cassini}}, \binits{G.D.}}:
1700,
{Sur des Taches du Soleil}.
\textit{Hist. Acad. Roy. Sci. Mém. Math. Phys.},
121.
\end{botherref}
\endbibitem

\bibitem[\protect\citeauthoryear{{Cassini}}{1730a}]{Cassini_Memoires_1730}
\begin{barticle}
\bauthor{\bsnm{{Cassini}}, \binits{G.D.}}:
\byear{1730}a,
\batitle{{D'ecouverte D'une Tache extraordinaire dans Jupiter, faite à
  l'Observatoire Royal}}.
\bjtitle{Mém. Acad. R. Sci., Depuis 1666 jusqu'a 1699}
\bvolume{10},
\bfpage{707}.
\end{barticle}
\endbibitem

\bibitem[\protect\citeauthoryear{{Cassini}}{1730b}]{Memoires_1730}
\begin{barticle}
\bauthor{\bsnm{{Cassini}}, \binits{G.D.}}:
\byear{1730}b,
\batitle{{Description d'une tache qui a paru dans le Soleil ce mois de May
  dernier 1684}}.
\bjtitle{Mém. Acad. R. Sci., Depuis 1666 jusqu'a 1699}
\bvolume{10},
\bfpage{653}.
\end{barticle}
\endbibitem

\bibitem[\protect\citeauthoryear{{Cassini}}{1730c}]{Cassini_Memoires_1730_Facula}
\begin{barticle}
\bauthor{\bsnm{{Cassini}}, \binits{G.D.}}:
\byear{1730}c,
\batitle{{Facules Observe'es Dans le Soleil le premier \& le second jour de
  Juin à l'Observatoire Royal, à la place de la Tache observée le mois de
  May, avec le retour de cette Tache à sa premiere forme}}.
\bjtitle{Mém. Acad. R. Sci., Depuis 1666 jusqu'a 1699}
\bvolume{10},
\bfpage{661}.
\end{barticle}
\endbibitem

\bibitem[\protect\citeauthoryear{{Cassini}}{1730d}]{Cassini_Memoires_1730_1688}
\begin{barticle}
\bauthor{\bsnm{{Cassini}}, \binits{G.D.}}:
\byear{1730}d,
\batitle{{Observation Des Tacher qui ont paru dans le Soleil le mois de May \&
  de Juin de 1688, avec une Méthode nouvelle de déterminer avec justesse la
  révolution du Soleil autour de son axe}}.
\bjtitle{Mém. Acad. R. Sci., Depuis 1666 jusqu'a 1699}
\bvolume{10},
\bfpage{727}.
\end{barticle}
\endbibitem

\bibitem[\protect\citeauthoryear{{Cassini} and
  {Maraldi}}{1701}]{Cassini_Maraldi_1701}
\begin{botherref}
\oauthor{\bsnm{{Cassini}}, \binits{G.D.}},
\oauthor{\bsnm{{Maraldi}}, \binits{J.P.}}:
1701,
{Tacher dans le Soleil observées le 29 Mars 1701}.
\textit{Hist. Acad. R. Sci. Mém. Math. Phys.},
78.
\end{botherref}
\endbibitem

\bibitem[\protect\citeauthoryear{{Cassini}}{1671}]{Cassini_1671_Ph_Tr}
\begin{barticle}
\bauthor{\bsnm{{Cassini}}, \binits{J.-D.}}:
\byear{1671},
\batitle{{New observations of spots in the Sun; made at the Royal Academy of
  Paris, the 11, 12 and 13th of August 1671; and English't out of the French,
  as follows}}.
\bjtitle{Philos. Trans. Royal Soc.}
\bvolume{6},
\bfpage{2250}.
\doiurl{https://doi.org/10.1098/rstl.1671.0042}.
\adsurl{https://doi.org/10.1098/rstl.1671.0042}.
\end{barticle}
\endbibitem

\bibitem[\protect\citeauthoryear{{Cassini}}{1701}]{Cassini_1701}
\begin{botherref}
\oauthor{\bsnm{{Cassini}}, \binits{J.}}:
1701,
{Des Tacher observées dans le Soleil au mois de Novembre de l'année 1700. au
  mois de Mars, \'{a} la fin d'Octoble, \& au mois Novembre cette année 1701}.
\textit{Hist. Acad. R. Sci. Mém. Math. Phys.},
262.
\end{botherref}
\endbibitem

\bibitem[\protect\citeauthoryear{{Cassini}}{1702a}]{Cassini_1702_2}
\begin{botherref}
\oauthor{\bsnm{{Cassini}}, \binits{J.}}:
1702a,
{Observation D'une nouvelle Tache dans le Soleil}.
\textit{Hist. Acad. R. Sci. Mém. Math. Phys.},
139.
\end{botherref}
\endbibitem

\bibitem[\protect\citeauthoryear{{Cassini}}{1702b}]{Cassini_1702}
\begin{botherref}
\oauthor{\bsnm{{Cassini}}, \binits{J.}}:
1702b,
{Observations De la Tache du Soleil, qui a paru le 6 Mai 1702}.
\textit{Hist. Acad. R. Sci. Mém. Math. Phys.},
131.
\end{botherref}
\endbibitem

\bibitem[\protect\citeauthoryear{{Cassini}}{1702\,--\,1704}]{Cassini_1702_1704}
\begin{bbook}
\bauthor{\bsnm{{Cassini}}, \binits{J.}}:
\byear{1702\,--\,1704},
\bbtitle{{Journal des observations faites à l'Observatoire de Paris et au
  château de Thury, 7 septembre 1702 - 16 mars 1704}},
\bpublisher{D3/21, Bibliothèque numérique},
\blocation{Observatoire de Paris, France}.
\end{bbook}
\endbibitem

\bibitem[\protect\citeauthoryear{{Cassini}}{1703}]{Cassini_1703}
\begin{botherref}
\oauthor{\bsnm{{Cassini}}, \binits{J.}}:
1703,
{Observation D'une Tache dans le Soleil}.
\textit{Hist. Acad. R. Sci. Mém. Math. Phys.},
15.
\end{botherref}
\endbibitem

\bibitem[\protect\citeauthoryear{{Clette} et~al.}{2023}]{2023SoPh..298...44C}
\begin{barticle}
\bauthor{\bsnm{{Clette}}, \binits{F.}},
\bauthor{\bsnm{{Lef{\`e}vre}}, \binits{L.}},
\bauthor{\bsnm{{Chatzistergos}}, \binits{T.}},
\bauthor{\bsnm{{Hayakawa}}, \binits{H.}},
\bauthor{\bsnm{{Carrasco}}, \binits{V.M.S.}},
\bauthor{\bsnm{{Arlt}}, \binits{R.}},
\bauthor{\bsnm{{Cliver}}, \binits{E.W.}},
\bauthor{\bsnm{{Dudok de Wit}}, \binits{T.}},
\bauthor{\bsnm{{Friedli}}, \binits{T.K.}},
\bauthor{\bsnm{{Karachik}}, \binits{N.}},
\bauthor{\bsnm{{Kopp}}, \binits{G.}},
\bauthor{\bsnm{{Lockwood}}, \binits{M.}},
\bauthor{\bsnm{{Mathieu}}, \binits{S.}},
\bauthor{\bsnm{{Mu{\~n}oz-Jaramillo}}, \binits{A.}},
\bauthor{\bsnm{{Owens}}, \binits{M.}},
\bauthor{\bsnm{{Pesnell}}, \binits{D.}},
\bauthor{\bsnm{{Pevtsov}}, \binits{A.}},
\bauthor{\bsnm{{Svalgaard}}, \binits{L.}},
\bauthor{\bsnm{{Usoskin}}, \binits{I.G.}},
\bauthor{\bsnm{{van Driel-Gesztelyi}}, \binits{L.}},
\bauthor{\bsnm{{Vaquero}}, \binits{J.M.}}:
\byear{2023},
\batitle{{Recalibration of the sunspot-number: status report}}.
\bjtitle{\solphys}
\bvolume{298},
\bfpage{44}.
\doiurl{https://doi.org/10.1007/s11207-023-02136-3}.
\adsurl{2023SoPh..298...44C}.
\end{barticle}
\endbibitem

\bibitem[\protect\citeauthoryear{{De La Lande}}{1778}]{La_Lande_1778}
\begin{botherref}
\oauthor{\bsnm{{De La Lande}}, \binits{J.J.L.}}:
1778,
{Second Mémoire sur les Taches du Soleil}.
\textit{Mém. Acad. R. Sci... avec les mémoires de mathématique \& de
  physique... tirez des registres de cette Académie},
393.
\end{botherref}
\endbibitem

\bibitem[\protect\citeauthoryear{{Ermolli} et~al.}{2023}]{2023ApJS..269...53E}
\begin{barticle}
\bauthor{\bsnm{{Ermolli}}, \binits{I.}},
\bauthor{\bsnm{{Chatzistergos}}, \binits{T.}},
\bauthor{\bsnm{{Giorgi}}, \binits{F.}},
\bauthor{\bsnm{{Carrasco}}, \binits{V.M.S.}},
\bauthor{\bsnm{{Aparicio}}, \binits{A.J.P.}},
\bauthor{\bsnm{{Chinnici}}, \binits{I.}}:
\byear{2023},
\batitle{{Solar observations by Angelo Secchi. I. digitization of original
  documents and analysis of group numbers over the period of 1853-1878}}.
\bjtitle{\apjs}
\bvolume{269},
\bfpage{53}.
\doiurl{https://doi.org/10.3847/1538-4365/ad0886}.
\adsurl{2023ApJS..269...53E}.
\end{barticle}
\endbibitem

\bibitem[\protect\citeauthoryear{{Flamsteed}}{1684}]{1684RSPT...14..535F}
\begin{barticle}
\bauthor{\bsnm{{Flamsteed}}, \binits{J.}}:
\byear{1684},
\batitle{{An account of a spot seen in the Sun from the 25th. of April to the
  8th. of May instant, with the line of its course predicted, if it make a
  second return}}.
\bjtitle{Phil. Trans. Roy. Soc. I}
\bvolume{14},
\bfpage{535}.
\adsurl{1684RSPT...14..535F}.
\end{barticle}
\endbibitem

\bibitem[\protect\citeauthoryear{{Flamsteedius}}{1712}]{Flamsteed_1712}
\begin{bbook}
\bauthor{\bsnm{{Flamsteedius}}, \binits{J.}}:
\byear{1712},
\bbtitle{{Historiae coelestis libri duo}},
\bpublisher{Typis J.~Matthews},
\blocation{Londini, UK}.
\end{bbook}
\endbibitem

\bibitem[\protect\citeauthoryear{{Hayakawa} et~al.}{2021}]{2021MNRAS.506..650H}
\begin{barticle}
\bauthor{\bsnm{{Hayakawa}}, \binits{H.}},
\bauthor{\bsnm{{Iju}}, \binits{T.}},
\bauthor{\bsnm{{Uneme}}, \binits{S.}},
\bauthor{\bsnm{{Besser}}, \binits{B.P.}},
\bauthor{\bsnm{{Kosaka}}, \binits{S.}},
\bauthor{\bsnm{{Imada}}, \binits{S.}}:
\byear{2021},
\batitle{{Reanalyses of the sunspot observations of Fogelius and Siverus: two
  'long-term' observers during the Maunder minimum}}.
\bjtitle{\mnras}
\bvolume{506},
\bfpage{650}.
\doiurl{https://doi.org/10.1093/mnras/staa2965}.
\adsurl{2021MNRAS.506..650H}.
\end{barticle}
\endbibitem

\bibitem[\protect\citeauthoryear{{Hayakawa} et~al.}{2022}]{2022ApJ...941..151H}
\begin{barticle}
\bauthor{\bsnm{{Hayakawa}}, \binits{H.}},
\bauthor{\bsnm{{Hattori}}, \binits{K.}},
\bauthor{\bsnm{{S{\^o}ma}}, \binits{M.}},
\bauthor{\bsnm{{Iju}}, \binits{T.}},
\bauthor{\bsnm{{Besser}}, \binits{B.P.}},
\bauthor{\bsnm{{Kosaka}}, \binits{S.}}:
\byear{2022},
\batitle{{An overview of sunspot observations in 1727-1748}}.
\bjtitle{\apj}
\bvolume{941},
\bfpage{151}.
\doiurl{https://doi.org/10.3847/1538-4357/ac6671}.
\adsurl{2022ApJ...941..151H}.
\end{barticle}
\endbibitem

\bibitem[\protect\citeauthoryear{{Hayakawa} et~al.}{2023}]{2023JSWSC..13...33H}
\begin{barticle}
\bauthor{\bsnm{{Hayakawa}}, \binits{H.}},
\bauthor{\bsnm{{Arlt}}, \binits{R.}},
\bauthor{\bsnm{{Iju}}, \binits{T.}},
\bauthor{\bsnm{{Besser}}, \binits{B.P.}}:
\byear{2023},
\batitle{{Karl von Lindener's sunspot observations during 1800-1827: another
  long-term dataset for the Dalton minimum}}.
\bjtitle{Journal of Space Weather and Space Climate}
\bvolume{13},
\bfpage{33}.
\doiurl{https://doi.org/10.1051/swsc/2023023}.
\adsurl{2023JSWSC..13...33H}.
\end{barticle}
\endbibitem

\bibitem[\protect\citeauthoryear{{Hayakawa}
  et~al.}{2024a}]{2024MNRAS.528.6280H}
\begin{barticle}
\bauthor{\bsnm{{Hayakawa}}, \binits{H.}},
\bauthor{\bsnm{{Carrasco}}, \binits{V.M.S.}},
\bauthor{\bsnm{{Aparicio}}, \binits{A.J.P.}},
\bauthor{\bsnm{{Villalba {\'A}lvarez}}, \binits{J.}},
\bauthor{\bsnm{{Vaquero}}, \binits{J.M.}}:
\byear{2024}a,
\batitle{{An overview of sunspot observations in the early Maunder minimum:
  1645-1659}}.
\bjtitle{\mnras}
\bvolume{528},
\bfpage{6280}.
\doiurl{https://doi.org/10.1093/mnras/stad3922}.
\adsurl{2024MNRAS.528.6280H}.
\end{barticle}
\endbibitem

\bibitem[\protect\citeauthoryear{{Hayakawa}
  et~al.}{2024b}]{2024ApJ...970L..31H}
\begin{barticle}
\bauthor{\bsnm{{Hayakawa}}, \binits{H.}},
\bauthor{\bsnm{{Murata}}, \binits{K.}},
\bauthor{\bsnm{{Teague}}, \binits{E.T.H.}},
\bauthor{\bsnm{{Bechet}}, \binits{S.}},
\bauthor{\bsnm{{S{\^o}ma}}, \binits{M.}}:
\byear{2024}b,
\batitle{{Analyses of Johannes Kepler's sunspot drawings in 1607: a revised
  scenario for the solar cycles in the early 17th century}}.
\bjtitle{\apjl}
\bvolume{970},
\bfpage{L31}.
\doiurl{https://doi.org/10.3847/2041-8213/ad57c9}.
\adsurl{2024ApJ...970L..31H}.
\end{barticle}
\endbibitem

\bibitem[\protect\citeauthoryear{{Henwood}, {Chapman}, and
  {Willis}}{2010}]{2010SoPh..262..299H}
\begin{barticle}
\bauthor{\bsnm{{Henwood}}, \binits{R.}},
\bauthor{\bsnm{{Chapman}}, \binits{S.C.}},
\bauthor{\bsnm{{Willis}}, \binits{D.M.}}:
\byear{2010},
\batitle{{Increasing lifetime of recurrent sunspot groups within the Greenwich
  photoheliographic results}}.
\bjtitle{\solphys}
\bvolume{262},
\bfpage{299}.
\doiurl{https://doi.org/10.1007/s11207-009-9419-5}.
\adsurl{2010SoPh..262..299H}.
\end{barticle}
\endbibitem

\bibitem[\protect\citeauthoryear{{Howard}, {Gilman}, and
  {Gilman}}{1984}]{1984ApJ...283..373H}
\begin{barticle}
\bauthor{\bsnm{{Howard}}, \binits{R.}},
\bauthor{\bsnm{{Gilman}}, \binits{P.I.}},
\bauthor{\bsnm{{Gilman}}, \binits{P.A.}}:
\byear{1984},
\batitle{{Rotation of the Sun measured from Mount Wilson white-light images}}.
\bjtitle{\apj}
\bvolume{283},
\bfpage{373}.
\doiurl{https://doi.org/10.1086/162315}.
\adsurl{1984ApJ...283..373H}.
\end{barticle}
\endbibitem

\bibitem[\protect\citeauthoryear{{Hoyt} and
  {Schatten}}{1998}]{1998SoPh..179..189H}
\begin{barticle}
\bauthor{\bsnm{{Hoyt}}, \binits{D.V.}},
\bauthor{\bsnm{{Schatten}}, \binits{K.H.}}:
\byear{1998},
\batitle{{Group sunspot numbers: a new solar activity reconstruction}}.
\bjtitle{\solphys}
\bvolume{179},
\bfpage{189}.
\doiurl{https://doi.org/10.1023/A:1005007527816}.
\adsurl{1998SoPh..179..189H}.
\end{barticle}
\endbibitem

\bibitem[\protect\citeauthoryear{{Husak} et~al.}{2023}]{2023SoPh..298..122H}
\begin{barticle}
\bauthor{\bsnm{{Husak}}, \binits{M.}},
\bauthor{\bsnm{{Braj{\v{s}}a}}, \binits{R.}},
\bauthor{\bsnm{{{\v{S}}poljari{\'c}}}, \binits{D.}},
\bauthor{\bsnm{{Krajnovi{\'c}}}, \binits{D.}},
\bauthor{\bsnm{{Ru{\v{z}}djak}}, \binits{D.}},
\bauthor{\bsnm{{Skoki{\'c}}}, \binits{I.}},
\bauthor{\bsnm{{Ro{\v{s}}a}}, \binits{D.}},
\bauthor{\bsnm{{Hr{\v{z}}ina}}, \binits{D.}}:
\byear{2023},
\batitle{{Bo{\v{s}}kovi{\'c}'s method for determining the axis and rate of
  solar rotation by observing sunspots in 1777}}.
\bjtitle{\solphys}
\bvolume{298},
\bfpage{122}.
\doiurl{https://doi.org/10.1007/s11207-023-02214-6}.
\adsurl{2023SoPh..298..122H}.
\end{barticle}
\endbibitem

\bibitem[\protect\citeauthoryear{{Illarionov} and
  {Arlt}}{2022}]{2022SoPh..297...79I}
\begin{barticle}
\bauthor{\bsnm{{Illarionov}}, \binits{E.}},
\bauthor{\bsnm{{Arlt}}, \binits{R.}}:
\byear{2022},
\batitle{{Reconstruction of the solar activity from the catalogs of the Zurich
  observatory}}.
\bjtitle{\solphys}
\bvolume{297},
\bfpage{79}.
\doiurl{https://doi.org/10.1007/s11207-022-02015-3}.
\adsurl{2022SoPh..297...79I}.
\end{barticle}
\endbibitem

\bibitem[\protect\citeauthoryear{{Illarionov} and
  {Arlt}}{2023}]{2023MNRAS.523.1809I}
\begin{barticle}
\bauthor{\bsnm{{Illarionov}}, \binits{E.}},
\bauthor{\bsnm{{Arlt}}, \binits{R.}}:
\byear{2023},
\batitle{{Sunspot positions from observations by Flaugergues in the Dalton
  minimum}}.
\bjtitle{\mnras}
\bvolume{523},
\bfpage{1809}.
\doiurl{https://doi.org/10.1093/mnras/stad1489}.
\adsurl{2023MNRAS.523.1809I}.
\end{barticle}
\endbibitem

\bibitem[\protect\citeauthoryear{{La Hire}}{1683\,--\,1684}]{La_Hire_1683_1684}
\begin{bbook}
\bauthor{\bsnm{{La Hire}}, \binits{P.}}:
\byear{1683\,--\,1684},
\bbtitle{{Journal des observations faites à l'Observatoire de Paris, 25
  janvier 1683 au 10 octobre 1684}},
\bpublisher{D2/1-10, Bibliothèque numérique},
\blocation{Observatoire de Paris, France}.
\end{bbook}
\endbibitem

\bibitem[\protect\citeauthoryear{{La Hire}}{1684\,--\,1686}]{La_Hire_1684_1686}
\begin{bbook}
\bauthor{\bsnm{{La Hire}}, \binits{P.}}:
\byear{1684\,--\,1686},
\bbtitle{{Journal des observations faites à l'Observatoire de Paris, 14
  octobre 1684 au 25 janvier 1686}},
\bpublisher{D2/1-10, Bibliothèque numérique},
\blocation{Observatoire de Paris, France}.
\end{bbook}
\endbibitem

\bibitem[\protect\citeauthoryear{{La Hire}}{1686\,--\,1687}]{La_Hire_1686_1687}
\begin{bbook}
\bauthor{\bsnm{{La Hire}}, \binits{P.}}:
\byear{1686\,--\,1687},
\bbtitle{{Journal des observations faites à l'Observatoire de Paris, 26
  janvier 1686 au 28 juin 1687}},
\bpublisher{D2/1-10, Bibliothèque numérique},
\blocation{Observatoire de Paris, France}.
\end{bbook}
\endbibitem

\bibitem[\protect\citeauthoryear{{La Hire}}{1687\,--\,1689}]{La_Hire_1687_1689}
\begin{bbook}
\bauthor{\bsnm{{La Hire}}, \binits{P.}}:
\byear{1687\,--\,1689},
\bbtitle{{Journal des observations faites à l'Observatoire de Paris, 1er
  juillet 1687 au 30 avril 1689}},
\bpublisher{D2/1-10, Bibliothèque numérique},
\blocation{Observatoire de Paris, France}.
\end{bbook}
\endbibitem

\bibitem[\protect\citeauthoryear{{La Hire}}{1689\,--\,1693}]{La_Hire_1689_1693}
\begin{bbook}
\bauthor{\bsnm{{La Hire}}, \binits{P.}}:
\byear{1689\,--\,1693},
\bbtitle{{Journal des observations faites à l'Observatoire de Paris, 1er
  juillet 1687 au 30 avril 1689}},
\bpublisher{D2/1-10, Bibliothèque numérique},
\blocation{Observatoire de Paris, France}.
\end{bbook}
\endbibitem

\bibitem[\protect\citeauthoryear{{La Hire}}{1693\,--\,1696}]{La_Hire_1693_1696}
\begin{bbook}
\bauthor{\bsnm{{La Hire}}, \binits{P.}}:
\byear{1693\,--\,1696},
\bbtitle{{Journal des observations faites à l'Observatoire de Paris, 14
  octobre 1693 au 20 d\'{e}cembre 1696}},
\bpublisher{D2/1-10, Bibliothèque numérique},
\blocation{Observatoire de Paris, France}.
\end{bbook}
\endbibitem

\bibitem[\protect\citeauthoryear{{La Hire}}{1697\,--\,1704}]{La_Hire_1697_1704}
\begin{bbook}
\bauthor{\bsnm{{La Hire}}, \binits{P.}}:
\byear{1697\,--\,1704},
\bbtitle{{Journal des observations faites à l'Observatoire de Paris, 4 janvier
  1697 au 20 juillet 1704}},
\bpublisher{D2/1-10, Bibliothèque numérique},
\blocation{Observatoire de Paris, France}.
\end{bbook}
\endbibitem

\bibitem[\protect\citeauthoryear{{La Hire}}{1700a}]{La_Hire_1700}
\begin{botherref}
\oauthor{\bsnm{{La Hire}}, \binits{P.}}:
1700a,
{Observation des Taches du Soliel qui ont paru au mois de Novembre 1700}.
\textit{Hist. Acad. Roy. Sci. Mém. Math. Phys.},
293.
\end{botherref}
\endbibitem

\bibitem[\protect\citeauthoryear{{La Hire}}{1700b}]{La_Hire_1700_2}
\begin{botherref}
\oauthor{\bsnm{{La Hire}}, \binits{P.}}:
1700b,
{Untitled note}.
\textit{Hist. Acad. Roy. Sci. Mém. Math. Phys.},
298.
\end{botherref}
\endbibitem

\bibitem[\protect\citeauthoryear{{La Hire}}{1701}]{La_Hire_1701}
\begin{botherref}
\oauthor{\bsnm{{La Hire}}, \binits{P.}}:
1701,
{Observation des Taches du Soliel qui ont paru vers les derniers jours du mois
  de Decembre de l'année dermiere 1700}.
\textit{Hist. Acad. Roy. Sci. Mém. Math. Phys.},
41.
\end{botherref}
\endbibitem

\bibitem[\protect\citeauthoryear{{La Hire}}{1702}]{La_Hire_1702}
\begin{botherref}
\oauthor{\bsnm{{La Hire}}, \binits{P.}}:
1702,
{Observation D'une Tache sur le Soleil, à Obsérvatoire}.
\textit{Hist. Acad. Roy. Sci. Mém. Math. Phys.},
137.
\end{botherref}
\endbibitem

\bibitem[\protect\citeauthoryear{{La Hire}}{1703}]{La_Hire_1703}
\begin{botherref}
\oauthor{\bsnm{{La Hire}}, \binits{P.}}:
1703,
{Observations D'une Tache, Qui a paru dans le Soleil au mois de Décembre 1702
  a l'Obsérvatoire}.
\textit{Hist. Acad. Roy. Sci. Mém. Math. Phys.},
16.
\end{botherref}
\endbibitem

\bibitem[\protect\citeauthoryear{{La Hire}}{1730}]{La_Hire_Memoires_1730}
\begin{barticle}
\bauthor{\bsnm{{La Hire}}, \binits{P.}}:
\byear{1730},
\batitle{{Observations d'une tache qui a paru sur le Disque du Soleil vers la
  fin du mois d'Avril \& au commencement de May de cette année 1686, faites à
  l'Observatoire}}.
\bjtitle{Mém. Acad. R. Sci., Depuis 1666 jusqu'a 1699}
\bvolume{10},
\bfpage{707}.
\end{barticle}
\endbibitem

\bibitem[\protect\citeauthoryear{{La Hire} and
  {Maraldi}}{1733}]{La_Hire_Hist_1695}
\begin{barticle}
\bauthor{\bsnm{{La Hire}}, \binits{P.d.}},
\bauthor{\bsnm{{Maraldi}}, \binits{J.P.}}:
\byear{1733},
\batitle{{Diverses observations Astromoniques}}.
\bjtitle{Mém. Acad. R. Sci., Depuis 1686 jusqu'a son Renouvellement 1699}
\bvolume{11},
\bfpage{264}.
\end{barticle}
\endbibitem

\bibitem[\protect\citeauthoryear{{Le Monnier}}{1741}]{Monnier_1741}
\begin{bbook}
\bauthor{\bsnm{{Le Monnier}}, \binits{P.-C.}}:
\byear{1741},
\bbtitle{{Histoire céleste, ou Recueil de toutes les observations
  astronomiques faites par ordre du Roy}},
\bpublisher{Chez Briasson},
\blocation{Paris, France}.
\end{bbook}
\endbibitem

\bibitem[\protect\citeauthoryear{{Lyu} et~al.}{2024}]{2024GSDJ...11..504L}
\begin{barticle}
\bauthor{\bsnm{{Lyu}}, \binits{Z.}},
\bauthor{\bsnm{{Ichikawa}}, \binits{K.}},
\bauthor{\bsnm{{Cheng}}, \binits{Y.}},
\bauthor{\bsnm{{Hayakawa}}, \binits{H.}},
\bauthor{\bsnm{{Kawamoto}}, \binits{Y.}}:
\byear{2024},
\batitle{{Digitization of weather records of Seungjeongwon Ilgi: a historical
  weather dynamics dataset of the Korean Peninsula in 1623{\textendash}1910}}.
\bjtitle{Geoscience Data Journal}
\bvolume{11},
\bfpage{504}.
\doiurl{https://doi.org/10.1002/gdj3.227}.
\adsurl{2024GSDJ...11..504L}.
\end{barticle}
\endbibitem

\bibitem[\protect\citeauthoryear{{Meeus}}{1991}]{1991aalg.book.....M}
\begin{bbook}
\bauthor{\bsnm{{Meeus}}, \binits{J.}}:
\byear{1991},
\bbtitle{{Astronomical algorithms}},
\bpublisher{Willmann-Bell},
\blocation{Richmind, Virginia}.
\adsurl{1991aalg.book.....M}.
\end{bbook}
\endbibitem

\bibitem[\protect\citeauthoryear{{Nagovitsyn}, {Pevtsov}, and
  {Osipova}}{2018}]{2018AstL...44..202N}
\begin{barticle}
\bauthor{\bsnm{{Nagovitsyn}}, \binits{Y.A.}},
\bauthor{\bsnm{{Pevtsov}}, \binits{A.A.}},
\bauthor{\bsnm{{Osipova}}, \binits{A.A.}}:
\byear{2018},
\batitle{{Two populations of sunspots: differential rotation}}.
\bjtitle{Astron. Lett.}
\bvolume{44},
\bfpage{202}.
\doiurl{https://doi.org/10.1134/S1063773718020056}.
\adsurl{2018AstL...44..202N}.
\end{barticle}
\endbibitem

\bibitem[\protect\citeauthoryear{{Neuh{\"a}user} and
  {Neuh{\"a}user}}{2021}]{2021AN....342..675N}
\begin{barticle}
\bauthor{\bsnm{{Neuh{\"a}user}}, \binits{R.}},
\bauthor{\bsnm{{Neuh{\"a}user}}, \binits{D.L.}}:
\byear{2021},
\batitle{{Critical comments on publications by S. Hoffmann and N. Vogt on
  historical novae/supernovae and their candidates}}.
\bjtitle{Astronomische Nachrichten}
\bvolume{342},
\bfpage{675}.
\doiurl{https://doi.org/10.1002/asna.202113872}.
\adsurl{2021AN....342..675N}.
\end{barticle}
\endbibitem

\bibitem[\protect\citeauthoryear{{Neuh{\"a}user}, {Arlt}, and
  {Richter}}{2018}]{2018AN....339..219N}
\begin{barticle}
\bauthor{\bsnm{{Neuh{\"a}user}}, \binits{R.}},
\bauthor{\bsnm{{Arlt}}, \binits{R.}},
\bauthor{\bsnm{{Richter}}, \binits{S.}}:
\byear{2018},
\batitle{{Reconstructed sunspot positions in the Maunder minimum based on the
  correspondence of Gottfried Kirch}}.
\bjtitle{Astronomische Nachrichten}
\bvolume{339},
\bfpage{219}.
\doiurl{https://doi.org/10.1002/asna.201813481}.
\adsurl{2018AN....339..219N}.
\end{barticle}
\endbibitem

\bibitem[\protect\citeauthoryear{{Newton} and
  {Nunn}}{1951}]{1951MNRAS.111..413N}
\begin{barticle}
\bauthor{\bsnm{{Newton}}, \binits{H.W.}},
\bauthor{\bsnm{{Nunn}}, \binits{M.L.}}:
\byear{1951},
\batitle{{The Sun's rotation derived from sunspots 1934-1944 and additional
  results}}.
\bjtitle{\mnras}
\bvolume{111},
\bfpage{413}.
\doiurl{https://doi.org/10.1093/mnras/111.4.413}.
\adsurl{1951MNRAS.111..413N}.
\end{barticle}
\endbibitem

\bibitem[\protect\citeauthoryear{{Ribes}, {Ribes}, and
  {Barthalot}}{1988}]{1988IAUS..123..227R}
\begin{bchapter}
\bauthor{\bsnm{{Ribes}}, \binits{E.}},
\bauthor{\bsnm{{Ribes}}, \binits{J.C.}},
\bauthor{\bsnm{{Barthalot}}, \binits{R.}}:
\byear{1988},
\bctitle{{Solar diameter and solar rotation during the Maunder minimum}}.
In: \beditor{\bsnm{{Christensen-Dalsgaard}}, \binits{J.}},
\beditor{\bsnm{{Frandsen}}, \binits{S.}} (eds.)
\bbtitle{Advances in Helio- and Asteroseismology},
\bsertitle{IAU Symposium}
\bseriesno{123},
\bfpage{227}.
\adsurl{1988IAUS..123..227R}.
\end{bchapter}
\endbibitem

\bibitem[\protect\citeauthoryear{{Ribes} and
  {Nesme-Ribes}}{1993}]{1993A&A...276..549R}
\begin{barticle}
\bauthor{\bsnm{{Ribes}}, \binits{J.C.}},
\bauthor{\bsnm{{Nesme-Ribes}}, \binits{E.}}:
\byear{1993},
\batitle{{The solar sunspot cycle in the Maunder minimum AD1645 to AD1715}}.
\bjtitle{\aap}
\bvolume{276},
\bfpage{549}.
\adsurl{1993A&A...276..549R}.
\end{barticle}
\endbibitem

\bibitem[\protect\citeauthoryear{{Schaefer}}{1991}]{1991QJRAS..32...35S}
\begin{barticle}
\bauthor{\bsnm{{Schaefer}}, \binits{B.E.}}:
\byear{1991},
\batitle{{Sunspot visibility}}.
\bjtitle{\qjras}
\bvolume{32},
\bfpage{35}.
\adsurl{1991QJRAS..32...35S}.
\end{barticle}
\endbibitem

\bibitem[\protect\citeauthoryear{{Schaefer}}{1993}]{1993ApJ...411..909S}
\begin{barticle}
\bauthor{\bsnm{{Schaefer}}, \binits{B.E.}}:
\byear{1993},
\batitle{{Visibility of sunspots}}.
\bjtitle{\apj}
\bvolume{411},
\bfpage{909}.
\doiurl{https://doi.org/10.1086/172895}.
\adsurl{1993ApJ...411..909S}.
\end{barticle}
\endbibitem

\bibitem[\protect\citeauthoryear{Spoerer}{1889}]{Spoerer1889}
\begin{bbook}
\bauthor{\bsnm{Spoerer}, \binits{G.}}:
\byear{1889},
\bbtitle{{Ueber die periodicitat der sonnenflecken seit dem Jahre 1618...}},
\bpublisher{Wilhelm Engelmann},
\blocation{Leipzig}.
\end{bbook}
\endbibitem

\bibitem[\protect\citeauthoryear{{Vokhmyanin} and
  {Zolotova}}{2023}]{2023SoPh..298..113V}
\begin{barticle}
\bauthor{\bsnm{{Vokhmyanin}}, \binits{M.}},
\bauthor{\bsnm{{Zolotova}}, \binits{N.}}:
\byear{2023},
\batitle{{Sunspot observations at the Eimmart observatory: revision and
  supplement}}.
\bjtitle{\solphys}
\bvolume{298},
\bfpage{113}.
\doiurl{https://doi.org/10.1007/s11207-023-02208-4}.
\adsurl{2023SoPh..298..113V}.
\end{barticle}
\endbibitem

\bibitem[\protect\citeauthoryear{{Vokhmyanin}, {Arlt}, and
  {Zolotova}}{2021}]{2021SoPh..296....4V}
\begin{barticle}
\bauthor{\bsnm{{Vokhmyanin}}, \binits{M.}},
\bauthor{\bsnm{{Arlt}}, \binits{R.}},
\bauthor{\bsnm{{Zolotova}}, \binits{N.}}:
\byear{2021},
\batitle{{Sunspot positions and areas from observations by Cigoli, Galilei,
  Cologna, Scheiner, and Colonna in 1612 - 1614}}.
\bjtitle{\solphys}
\bvolume{296},
\bfpage{4}.
\doiurl{https://doi.org/10.1007/s11207-020-01752-7}.
\adsurl{2021SoPh..296....4V}.
\end{barticle}
\endbibitem

\bibitem[\protect\citeauthoryear{{Wagner}, {Neuh{\"a}user}, and
  {Arlt}}{2023}]{2023AN....34420078W}
\begin{barticle}
\bauthor{\bsnm{{Wagner}}, \binits{D.}},
\bauthor{\bsnm{{Neuh{\"a}user}}, \binits{R.}},
\bauthor{\bsnm{{Arlt}}, \binits{R.}}:
\byear{2023},
\batitle{{Auroral oval reconstruction for historical geomagnetic storms in the
  18th and 19th century}}.
\bjtitle{Astronomische Nachrichten}
\bvolume{344},
\bfpage{e20220078}.
\doiurl{https://doi.org/10.1002/asna.20220078}.
\adsurl{2023AN....34420078W}.
\end{barticle}
\endbibitem

\bibitem[\protect\citeauthoryear{{Wang} and {Li}}{2022}]{2022SoPh..297..127W}
\begin{barticle}
\bauthor{\bsnm{{Wang}}, \binits{H.}},
\bauthor{\bsnm{{Li}}, \binits{H.}}:
\byear{2022},
\batitle{{Rediscovery of 23 historical records of naked-eye sunspot
  observations in AD 1618}}.
\bjtitle{\solphys}
\bvolume{297},
\bfpage{127}.
\doiurl{https://doi.org/10.1007/s11207-022-02046-w}.
\adsurl{2022SoPh..297..127W}.
\end{barticle}
\endbibitem

\bibitem[\protect\citeauthoryear{{Zolotova} and
  {Vokhmyanin}}{2025}]{2025SoPh..300...17Z}
\begin{barticle}
\bauthor{\bsnm{{Zolotova}}, \binits{N.}},
\bauthor{\bsnm{{Vokhmyanin}}, \binits{M.}}:
\byear{2025},
\batitle{{Long-lived sunspots in historical records: a case study analysis from
  1660 to 1676}}.
\bjtitle{\solphys}
\bvolume{300},
\bfpage{17}.
\doiurl{https://doi.org/10.1007/s11207-025-02432-0}.
\adsurl{2025SoPh..300...17Z}.
\end{barticle}
\endbibitem

\end{thebibliography}

\IfFileExists{\jobname.bbl}{} {\typeout{}
\typeout{****************************************************}
\typeout{****************************************************}
\typeout{** Please run "bibtex \jobname" to obtain} \typeout{**
the bibliography and then re-run LaTeX} \typeout{** twice to fix
the references !}
\typeout{****************************************************}
\typeout{****************************************************}
\typeout{}}

\end{document}